\documentclass[10pt, journal,twocolumn]{IEEEtran}
\IEEEoverridecommandlockouts
\usepackage{amsthm, amsmath,amssymb,amsfonts, graphicx,textcomp,xcolor,xpatch, tikz, csquotes, epstopdf,flowchart, pgf, bmpsize, adjustbox, pdftexcmds,pgfplots,  floatrow,float, array, arydshln, multirow, rotating, bm, amsbsy, pst-node, lipsum, tikzscale, tkz-euclide,mathrsfs,booktabs, mathtools, makecell, float}
\usepackage[utf8]{inputenc}
\usepackage[edges]{forest}
\usetikzlibrary{positioning}
\usepackage[english]{babel}
\usepackage[linesnumbered,ruled,vlined]{algorithm2e}
\usepackage[draft]{hyperref}
\usepackage[backend=bibtex,bibstyle=ieee,citestyle=numeric-comp,url=false,doi=false,isbn=false]{biblatex}
\bibliography{SequentialDecoding.bib}
\def\BibTeX{{\rm B\kern-.05em{\sc i\kern-.025em b}\kern-.08em
    T\kern-.1667em\lower.7ex\hbox{E}\kern-.125emX}}

\hypersetup{
  colorlinks   = true, 
  urlcolor     = blue, 
  linkcolor    = blue, 
  citecolor   = red 
}

\newcommand{\eref}[1]{(\ref{#1})}               

\usepackage{pgfplots}\pgfplotsset{compat=newest}

\allowdisplaybreaks

\tikzset{
	dot/.style={circle,draw,inner sep=1.2,fill=black},
}

\makeatletter
\xpatchcmd{\proof}{\topsep6\p@\@plus6\p@\relax}{}{}{}
\makeatother
\newcounter{example}[section]

\newtheorem{rem}{Remark}

\graphicspath{{./figures/}}

\floatstyle{plaintop}
\restylefloat{table}

    \newcommand\Labeliii{%
    	node[below,midway,yshift=1pt,font=\scriptsize]{$1$}%
    }
    \newcommand\Labeliv{%
    	node[above,midway,yshift=-1.5pt,font=\scriptsize]{$0$}%
    }

\SetCommentSty{mycommfont}

\SetKwInput{KwInput}{Input}                
\SetKwInput{KwOutput}{Output}              
\SetKwRepeat{Do}{do}{while}%
\SetKwProg{Init}{init}{}{}

\newcommand{\bfd}{{\boldsymbol d}}

\newcommand{\bfs}{{\boldsymbol s}}

\newcommand{\bfv}{{\boldsymbol v}}

\newcommand{\bfx}{{\boldsymbol x}}
\newcommand{\bfy}{{\boldsymbol y}}

\renewcommand{\baselinestretch}{0.985}
\expandafter\def\expandafter\normalsize\expandafter{%
    \normalsize%
    \setlength\abovedisplayskip{3.9pt}%
    \setlength\belowdisplayskip{3.9pt}%
    \setlength\abovedisplayshortskip{-0.5pt}%
    \setlength\belowdisplayshortskip{-0.5pt}%
}

\begin{document}

\title{Sequential Decoding of Multiple Sequences for Synchronization Errors 
\thanks{This work has been supported by the European Research
		Council (ERC) under the European Union’s Horizon 2020 research and innovation programme (Grant Agreement No. 801434).
		
		The authors are with the Institute for Communications Engineering, Technical University of Munich, DE-80333 Munich, Germany,
 		Emails: anisha.banerjee@tum.de, andreas.lenz@mytum.de, antonia.wachter-zeh@tum.de 
 		
 		A part of this work was presented at \emph{2022 IEEE Information Theory Workshop (ITW)} \cite{banerjeeSequential2022}. }  } 

\author{%
	\IEEEauthorblockN{Anisha Banerjee, Andreas Lenz and Antonia Wachter-Zeh}
}

\maketitle

\begin{abstract}
Sequential decoding, commonly applied to substitution channels, is a sub-optimal alternative to Viterbi decoding with significantly reduced memory costs.
	In this work, a sequential decoder for convolutional codes over channels that are prone to insertion, deletion, and substitution errors, is described and analyzed. 
	Our decoder expands the code trellis by a new channel-state variable, called drift state, as proposed by Davey and MacKay. A suitable decoding metric on that trellis for sequential decoding is derived, generalizing the original Fano metric. The decoder is also extended to facilitate the simultaneous decoding of multiple received sequences that arise from a single transmitted sequence. Under low-noise environments, our decoding approach reduces the decoding complexity by a couple orders of magnitude in comparison to Viterbi's algorithm, albeit at slightly higher bit error rates. An analytical method to determine the computational cutoff rate is also suggested. This analysis is supported with numerical evaluations of bit error rates and computational complexity, which are compared with respect to optimal Viterbi decoding.
\end{abstract}

\section{Introduction}

Most error-control systems operate under the assumption of perfect synchronization between transmitter and receiver, while their respective decoders are designed to detect and correct substitution errors alone. However, when this assumption does not hold, as in the case of some networking and data storage channels \cite{yazdi_dna-based_2015,heckel_characterization_2019}, some transmitted symbols may be lost or random ones may be inserted into the received stream. Such errors are referred to as deletions and insertions.

There exists rich literature dedicated to the study of channels that are susceptible to insertion, deletion and substitution errors, and suitable error-correcting codes to increase transmission reliability under such environments \cite{levenshtein_binary_1966, mercier_survey_2010, davey_reliable_2001, calabi_general_1969, coumou_insertion_2008, briffa_timevarying_2014, gallager_sequential_1961, rahmati_bounds_2013-1}. In this work, we are interested in the use of convolutional codes for the purpose of correcting these errors. Prior work in \cite{mansour_convolutional_2010, buttigieg_improved_2015} suggested new trellis structures that helped in adapting the conventional Viterbi and MAP decoders to handle insertions and deletions. Concatenated schemes \cite{maaroufConcatenatedCodesMultiple2023} with inner convolutional codes have built on these trellis structures to decode from multiple sequences over insertion and deletion channels. One drawback of these decoding approaches however, lies in their memory requirements and computational complexity. In particular, the trellis grows rapidly with factors like constraint length of the code, number of information blocks per codeword and maximum allowable insertions and deletions per block. This motivated us to look into an alternative approach, namely \emph{sequential decoding} with the goal to achieve a low-complexity decoding approach. 

First proposed by Wozencraft \cite{wozencraft_sequential_1957}, sequential decoders constitute a decoding strategy that is suited to convolutional codes with high constraint lengths. This is primarily owed to the fact that a typical sequential decoder will only examine those codewords that seem likely to have been transmitted, unlike the Viterbi decoder which assesses all possibilities, regardless of noise levels. Although this leads to a worse error-correcting performance in sequential decoding compared to Viterbi decoding, the resulting decoding complexity is effectively independent of the encoder's memory.

The main objective of this work is to tailor the sequential decoding approach for use in channels that experience insertions, deletions as well as substitution errors. This problem was first addressed by Gallager \cite{gallager_sequential_1961}, who used Wozencraft's original algorithm to implement the sequential decoder, and subsequently analyzed its complexity. Mansour and Tewfik \cite{mansour_convolutional_2002} also worked on this problem, by adopting a new trellis structure and specifically limiting their focus to the stack algorithm. However, unlike \cite{gallager_sequential_1961, mansour_convolutional_2002}, this work formulates a new decoding metric, wherein the likelihood component is computed using the lattice metric \cite{bahl_decoding_1975}, and an additional bias term accounts for probability of the predicted message sequence and that of the received bit sequence. Furthermore, we employ the trellis structure proposed in \cite{buttigieg_improved_2015} and limit our attention to Fano's algorithm, which typically performs fewer computations and explores more paths compared to other variants of sequential decoding. Using the approach in \cite{johannesson_fundamentals_2015}, an analytical assessment of the algorithm's average complexity is also performed.

Since DNA storage typically involves the synthesis of data into numerous short DNA strands, it is also crucial to study the problem of recovering the stored data from multiple erroneous copies of the original strand. In this regard, \cite{maaroufConcatenatedCodesMultiple2023} proposed concatenated coding schemes for transmitting a single DNA sequence over multiple parallel channels and also decoding algorithms for multiple received sequences. The authors of \cite{srinivasavaradhanTrellis2021} also suggested a sub-optimal decoding algorithm of lower complexity for multiple sequences. Likewise, we also extend our sequential decoding framework to enable the simultaneous decoding of multiple received sequences.

The structure of this paper is as follows. We start by briefly considering the modeling of channels which are vulnerable to insertions, deletions and substitution errors, followed by a discussion on the decoding algorithm in Section~\ref{sec::prelim}. Following this, Section~\ref{sec:decoder:metric} demonstrates how we adapt sequential decoders to our specific channel model by deriving a new decoding metric that suitably generalizes the original metric proposed by Fano \cite{fano_heuristic_1963}. An asymptotic approximation of this decoding metric is also derived to assist in the analytical assessment of the average computational complexity of this decoder, which is detailed in Section~\ref{sec::comp_an}. Additionally in Section~\ref{sec::multiple}, we suggest how this decoder can be extended to permit the simultaneous decoding of multiple sequences, while also repeating the computational analysis for the same. Finally, Section~\ref{sec::results} presents simulation results that compare the performance of our decoder to the Viterbi decoder, specifically in regard to bit error rates and computational effort.

\section{Preliminaries} \label{sec::prelim}
\subsection{Channel model}
\begin{figure}[]
	\centering
	\begin{tikzpicture}[->,>=stealth',auto,node distance=3.5cm,scale=0.9, every node/.style={scale=0.9}]
	\node[name=a, circle, inner sep=0pt, draw=black,fill=white, minimum size=1cm] at (0,-2){$x_{i}$};
	\node[name=b, circle, inner sep=0pt, draw=black,fill=white, minimum size=1cm] at (6.0,-2){$x_{i+1}$};
	\node[anchor=east,xshift=-15pt] at (a) {$\cdots$};
	\node[anchor=east,xshift=40pt] at (b) {$\cdots$};
	\path[]
	(a) edge[bend right=35, dashed] node [midway, below, yshift=-.7mm] {$P_d$} node[midway, above] {Delete}(b)
	(a) edge[bend left=35] node [midway, above, right, yshift=3mm, xshift=-18.5mm,name=tr] {Transmit with no error} node[midway, below] {$P_t(1-P_s)$} (b);
	\draw (a) edge[loop above, out=120, in=60,min distance=20mm] node {Insert} node[midway, left, xshift=-5mm, yshift=-8mm] {$P_i$}(a);
	\draw (a) ->node [above] {Transmit with substitution} node [below] {$P_tP_s$}  (b);
	\end{tikzpicture}
	\caption{Allowed transitions in the state machine model for the insertion, deletion, and substitution channel \cite{davey_reliable_2001}}
	\label{fig::si}
\end{figure}
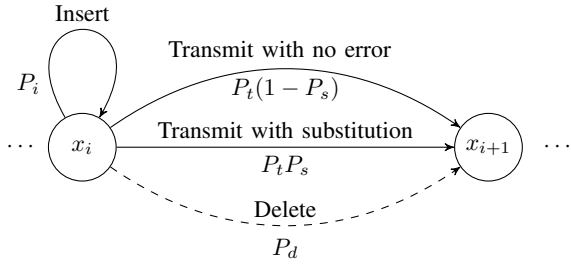

As in \cite{davey_reliable_2001,bahl_decoding_1975}, we adopt a finite-state-machine model for our channel, specified by three parameters $P_i$, $P_d$ and $P_s$, which denote the insertion, deletion and substitution probabilities of the channel, respectively. Let $\boldsymbol{x}_1^R=(x_1, \ldots x_R) \in \{0,1\}^R$ denote a sequence of bits awaiting transmission. 
From Fig.~\ref{fig::si}, we observe that under this construct, for each input bit, one of four events may occur: a random bit is inserted into the received stream with probability $P_i$ and $x_i$ remains in the transmission queue; or the next bit queued for transmission, i.e., $x_i$, is deleted with probability $P_d$; or $x_i$ is received at output end, either erroneously or correctly, with probabilities $P_tP_s$ and $P_t(1-P_s)$ respectively. Here, $P_t=1-P_i-P_d$ simply refers to the transmission probability (possibly with a substitution error).
\subsection{Convolutional codes}
Before describing the decoding framework, we shortly recapitulate the basics of convolutional codes. These codes constitute a special category of tree codes, that incorporate memory and aim to encode a stream of input bits in a block-wise manner, by means of shift registers. They are typically specified by three parameters: $[c,b,m]$, indicating that for every $b$ input bits received, the encoder generates $c$ output bits, which are a function of the last $b(m+1)$ input bits. 

\begin{figure}[!htb]
	\centering
	\begin{tikzpicture}[>=latex',font={\sf \small}, transform shape, squarednode/.style={rectangle,draw, minimum size=7mm}, roundnode/.style={circle, draw, minimum size=6mm}, dot/.style={circle,draw,inner sep=1.2,fill=black}]
	\node[squarednode] (a) {};
	\node[roundnode, right=1cm of a] (d1) {+};
	\node[dot, left=1cm of a] (d4) {};
	\node[left=0.5cm of d4] (d5) {$\boldsymbol{u}$};
	\node[dot, above=1cm of d1] (p1){};
	\node[roundnode, below=1cm of d1] (p2) {+};
	\node[right=0.75cm of p1.center] (o1) {};
	\node[right=0.75cm of p2.center] (o3) {};
	\node[right=0.75cm of d1.center] (o2) {};
	\draw[->] (p1) -- (d1);
	\draw[->] (d4) |- (p1);
	\draw[->] (d4) |- (p2);
	\draw[->] (d5) -- (a);
	\draw[->] (a) -- (d1) node [midway, dot, name=ad1] {};
	\draw[->] (p1) -- (o1);
	\draw[->] (p2) -- (o3);
	\draw[->] (d1) -- (o2);
	\draw[->] (ad1) |- ($(ad1)+(0, -0.8)$) -- (p2);
	\node[xshift=5pt, yshift=2pt] at (o1) {$\boldsymbol{x}^{(1)}$};
	\node[xshift=5pt, yshift=2pt] at (o2) {$\boldsymbol{x}^{(2)}$};
	\node[xshift=5pt, yshift=2pt] at (o3) {$\boldsymbol{x}^{(3)}$};
	\end{tikzpicture}
	\caption{A $[3,1,1]$ binary convolutional encoder}
	\label{fig::ccode}
\end{figure}
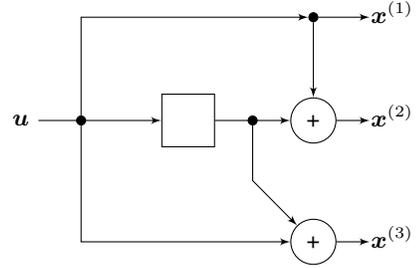

To exemplify this, Fig. \ref{fig::ccode} depicts the encoder for a binary $[3,1,1]$ convolutional code. It accepts $b=1$ input, has $c=3$ outputs, and includes $m=1$ memory unit, the contents of which indicate the \emph{encoder state}. Thus, a total of two states are possible. 
\subsection{Joint code and channel tree structure}
As in \cite{buttigieg_improved_2015}, the vector obtained at the receiving end of the channel is viewed as the output of a hidden Markov model (HMM), where each hidden state is a pair of the encoder state and the drift value. In this context, we define the \emph{drift}~\cite{davey_reliable_2001} as the difference between number of bits received and transmitted. In particular, $d_i$ is used to signify the net drift accumulated after the transmission of $i$ bits. 
For more details about the drift variable, we refer the reader to \cite{davey_reliable_2001}.

The decoder works on a tree representation of this HMM, such that any path in this tree describes how the encoder state and net drift value could change over time. As a demonstration, Fig. \ref{fig::hmm_code} depicts the joint code and channel tree for a $[3,1,1]$ convolutional encoder. For any given sequence of HMM states, the concatenation of edge labels along the respective path in the code tree indicates the originally transmitted codeword.
\subsection{Fano's Algorithm}
In this work, we limit our focus to a particular variant of sequential decoding, namely Fano's algorithm \cite{fano_heuristic_1963}. It employs a tree structure as described earlier, and assigns a metric\footnote{Discussed further in Section~\ref{sec:decoder:metric}.} to each node of this tree to suitably quantify the closeness of the received sequence to the sequence predicted by that node. The algorithm operates on the principle that node metrics along the correct path keep increasing on an average. 
The algorithm searches for such a path, by tracking metrics of the current, previous and best successor nodes, denoted as $\mu_c$, $\mu_p$ and $\mu_s$ respectively. A dynamic threshold $T$ is used to check for the aforementioned property. This variable can only be altered by integer multiples of a user-defined step size $\Delta$. Starting from the tree root, the decoder works as follows:
\begin{figure}[]
	\centering
	\scalebox{0.9}{
		\begin{forest}
			[ ,name=a,for tree={s sep=3pt,l sep=3cm,dot,grow=0}
			[, name=b1, edge label={node[below,midway,yshift=-3pt,font=\scriptsize]{$111$}},
			[, name=c1, edge label={node[below,midway,yshift=1pt,font=\scriptsize]{$100$}}]
			[, name=c2]
			[, name=c3]
			[, name=c4, edge={dashed}]
			[, name=c5, edge={dashed}]
			[, name=c6, edge label={node[above,midway,yshift=-1.5pt,font=\scriptsize]{$011$}}, edge={dashed}]]
			[, name=b2, edge label={node[below,midway,yshift=-1pt,font=\scriptsize]{$111$}},
			[, name=c7, edge label={node[below,midway,yshift=1pt,font=\scriptsize]{$100$}}]
			[, name=c8]
			[, name=c9]
			[, name=c10, edge={dashed}]
			[, name=c11, edge={dashed}]
			[, name=c12, edge label={node[above,midway,yshift=-1.5pt,font=\scriptsize]{$011$}}, edge={dashed}]]
			[, name=b3, edge label={node[below,midway,yshift=1pt,font=\scriptsize]{$111$}},
			[, name=c13, edge label={node[below,midway,yshift=1pt,font=\scriptsize]{$100$}}]
			[, name=c14]
			[, name=c15]
			[, name=c16, edge={dashed}]
			[, name=c17, edge={dashed}]
			[, name=c18, edge label={node[above,midway,yshift=-1.5pt,font=\scriptsize]{$011$}}, edge={dashed}]]
			[, name=b4, edge label={node[above,midway,yshift=-1.5pt,font=\scriptsize]{$000$}}, edge={dashed},
			[, name=c19, edge label={node[below,midway,yshift=1pt,font=\scriptsize]{$111$}}]
			[, name=c20]
			[, name=c21]
			[, name=c22, edge={dashed}]
			[, name=c23, edge={dashed}]
			[, name=c24, edge label={node[above,midway,yshift=-1.5pt,font=\scriptsize]{$000$}}, edge={dashed}]]
			[, name=b5, edge label={node[above,midway,yshift=1.5pt,font=\scriptsize]{$000$}}, edge={dashed},
			[, name=c25, edge label={node[below,midway,yshift=1pt,font=\scriptsize]{$111$}}]
			[, name=c26]
			[, name=c27]
			[, name=c28, edge={dashed}]
			[, name=c29, edge={dashed}]
			[, name=c30, edge label={node[above,midway,yshift=-1.5pt,font=\scriptsize]{$000$}}, edge={dashed}]]
			[, name=b6, edge label={node[above,midway,yshift=3.5pt,font=\scriptsize]{$000$}}, edge={dashed},
			[, name=c31, edge label={node[below,midway,yshift=1pt,font=\scriptsize]{$111$}}]
			[, name=c32]
			[, name=c33]
			[, name=c34, edge={dashed}]
			[, name=c35, edge={dashed}]
			[, name=c36, edge label={node[above,midway,yshift=-1.5pt,font=\scriptsize]{$000$}}, edge={dashed}]]]
			\node[yshift=-0.28cm,xshift=-0.3cm,font=\footnotesize] at (a) {$(S_0, 0)$};
			\node[yshift=-0.25cm,font=\footnotesize] at (b1) {$(S_1, 1)$};
			\node[yshift=-0.25cm,font=\footnotesize] at (b2) {$(S_1, 0)$};
			\node[yshift=-0.25cm,font=\footnotesize] at (b3) {$(S_1, -1)$};
			\node[yshift=0.25cm,font=\footnotesize] at (b4) {$(S_0, 1)$};
			\node[yshift=0.25cm,font=\footnotesize] at (b5) {$(S_0, 0)$};
			\node[yshift=0.25cm,font=\footnotesize] at (b6) {$(S_0, -1)$};
			\node[xshift=0.5cm,font=\footnotesize] at (c1) {$(S_1, 2)$};
			\node[xshift=0.6cm,font=\footnotesize] at (c36) {$(S_0, -2)$};
	\end{forest}}
	\caption{Joint code and channel tree of a $[3,1,1]$ convolutional code. Dashed lines correspond to an input of $0$ and solid lines to input $1$. Each node has two state variables, the convolutional code state and the  drift state.}
	\label{fig::hmm_code}
\end{figure}
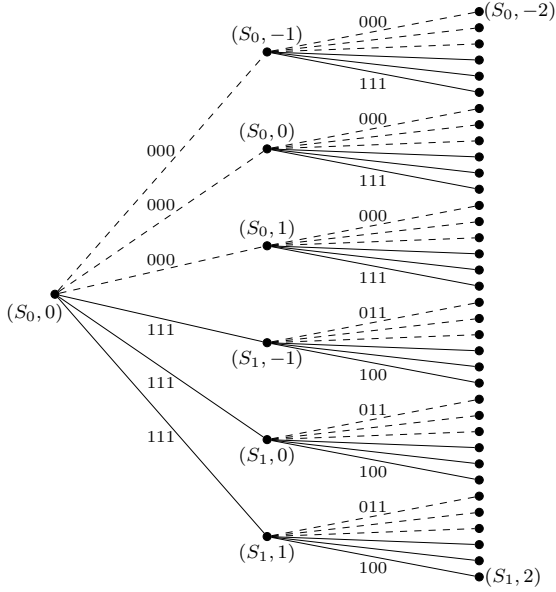
\begin{enumerate}
	\item If best successor has metric $\mu_s \geq T$, move forward.
	\begin{itemize}
		\item If this node has never been visited before, $T$ is tightened such that
	\end{itemize}
\begin{equation}
	T\leq \mu_c < T+\Delta. \label{eq::fv}
\end{equation}
	\item Else, step back to the immediate predecessor.
	\begin{itemize}
		\item If other successors with metrics above $T$ exist, the decoder steps forward to it, as in step 1).
		\item Else: \begin{itemize}
			\item If $\mu_p < T$, lower $T$ by $\Delta$.
			\item Else repeat step 2).
		\end{itemize}
	\end{itemize}
\end{enumerate} 
In this manner, all the paths with metrics above or equal to the threshold $T$ are systematically explored.
\section{Decoder metric} \label{sec:decoder:metric}
From the preceding discussion, it is evident that a metric for each tree node  must be defined to quantify its closeness to the received vector. Specifically, this metric should help minimize the probability of choosing a wrong successor.

\subsection{Metric Definition Based on Posterior Distribution}

Let $\boldsymbol{y}_1^{N}=(y_1, \ldots y_{N})$ denote a received sequence produced by the transmission of a codeword of $L$ blocks. For a convolutional code with parameters $[c,b]$, consider a node at depth $t$, say $\bfv_0^t$, that is reached from the root via the convolutional code states $\boldsymbol{s}_0^{t}=(s_0, \ldots, s_t)$ and the drift states $\boldsymbol{d}_0^{t}=(d_0, d_c,\ldots d_{ct})$\footnote{The time indices of the drift values are integer multiples of $c$, since a node at depth $i$ is reached after the transmission of $ci$ bits.}, where initial drift is $d_0=0$. We may represent this node in vector form as 
\begin{equation*}
    \bfv_0^t=((s_0,d_0), (s_1,d_c), \ldots, (s_t,d_{ct})).
\end{equation*}    
Then, we define the metric of node $\boldsymbol{v}_0^t$ to be
\begin{align}
\mu(\boldsymbol{v}_0^t)
&= \log_2 P(\boldsymbol{v}_0^t, \boldsymbol{y}_1^N) -\log_2 P(\boldsymbol{y}_1^N).  \label{eq::fanoL}
\end{align}

Hence, the decoder metric of a tree node is essentially the probability of its predicted codeword and drift changes, given a specific received frame. This definition is in the same spirit as that in \cite{massey_variable-length_1972}, wherein Massey proved the optimality of the Fano metric in the context of binary symmetric channels.

Before further simplifying \eref{eq::fanoL}, we recognize that the path traced from the root by $\boldsymbol{v}_0^t$, only accounts for the first $ct+d_{ct}$ symbols of the received vector. The remaining symbols $\boldsymbol{y}_{ct+d_{ct}+1}^N$, are assumed to have been produced by a tailing message sequence that guides the convolutional encoder through the states $\tilde{\boldsymbol{s}}_{t+1}^L=(s_{t+1}, \ldots, s_L)$. We additionally make the simplifying assumption that $\boldsymbol{y}_{ct+d_{ct}+1}^N$ is unaffected by bits transmitted prior to it\footnote{Exactly true for bits $\boldsymbol{y}^N_{c(t+m)+d_{c(t+m)}+1}$.} and we may write
\begin{align}
P(\boldsymbol{v}_0^t, \tilde{\boldsymbol{s}}_{t+1}^L, \boldsymbol{y}_1^N)
 &=P(\boldsymbol{v}_0^t, \bfy_1^{ct+d_{ct}}) P(\bfy_{ct+d_{ct}+1}^{N}, \tilde{\bfs}_{t+1}^L | \bfv_0^t) \nonumber
 \\
&=  P(\boldsymbol{v}_0^t,\boldsymbol{y}_1^{ct+d_{ct}}) P(\tilde{\boldsymbol{s}}_{t+1}^L, \boldsymbol{y}_{ct+d_{ct}+1}^N| s_t, d_{ct}). \nonumber
\end{align}
Marginalizing this term over all possible message tails,
\begin{align}
P(\boldsymbol{v}_0^t&,\boldsymbol{y}_1^N)=\sum_{ \tilde{\boldsymbol{s}}_{t+1}^L } P(\boldsymbol{v}_0^t, \tilde{\boldsymbol{s}}_{t+1}^L, \boldsymbol{y}_1^N) \nonumber \\
&=  P(\boldsymbol{v}_0^t,\boldsymbol{y}_1^{ct+d_{ct}})  \sum_{ \tilde{\boldsymbol{s}}_{t+1}^L }  P( \tilde{\boldsymbol{s}}_{t+1}^L,\boldsymbol{y}_{ct+d_{ct}+1}^N| s_t, d_{ct}).  \label{eq::fano_jp}
\end{align}

Both \eref{eq::fanoL} and \eref{eq::fano_jp} require us to evaluate the probability of receiving a particular sequence. To make the dependence of this quantity on the length of the causal transmitted sequence more explicit, we introduce the following notation.
\begin{equation}
	P_R(\boldsymbol{y}_1^N)=\sum_{ \boldsymbol{x} \in \{0,1\}^{R}} P(\boldsymbol{x}, \boldsymbol{y}_1^N). \label{eq::pt}
\end{equation}

Applying \eref{eq::fano_jp} and \eref{eq::pt} in \eref{eq::fanoL}, we arrive at the following definition of decoder metric,
\begin{align}
\mu(\boldsymbol{v}_0^t)=&~ \log_2P(\boldsymbol{v}_0^t, \boldsymbol{y}_1^{ct+d_{ct}})+\log_2P_{c(L-t)}(\boldsymbol{y}_{ct+d_{ct}+1}^N|d_{ct})\nonumber \\
&-\log_2P_{cL}(\boldsymbol{y}_1^N) \nonumber \\
=&~ \log_2 P(\boldsymbol{s}_0^{t})+ \log_2 P(\boldsymbol{d}_0^{t})+\log_2 P(\boldsymbol{y}_1^{ct+d_{ct}}|\boldsymbol{v}_0^t)\nonumber \\
&+\log_2P_{c(L-t)}(\boldsymbol{y}_{ct+d_{ct}+1}^N|d_{ct}) -\log_2P_{cL}(\boldsymbol{y}_1^N)\nonumber\\
=&~ \! \sum_{i=0}^{t-1}\log_2 P(s_{i+1}|s_i)+\log_2 P(d_0)+\! \sum_{i=0}^{t-1}\! \pi_i +\sum_{i=0}^{t-1}\gamma_i(\boldsymbol{y}) \nonumber \\ &+ \log_2 \frac{P_{c(L-t)}(\boldsymbol{y}_{ct+d_{ct}+1}^N|d_{ct})}{P_{cL}(\boldsymbol{y}_1^N)},  \label{eq::fano_final} 
\end{align}
where $\pi_i$ indicates the logarithmic drift transition probability of the $i^{\mathrm{th}}$ branch, while $\gamma_i(\boldsymbol{y})$ signifies the likelihood of the received bits given the state and drift transitions specified by the $i^{\mathrm{th}}$ branch. More explicitly, we define these variables as
\begin{align}
    \pi_i&=\log_2 P(d_{c(i+1)}|d_{ci}) \nonumber \\
    \gamma_i(\boldsymbol{y})&=\log_2 P(\boldsymbol{y}_{ci+d_{ci}+1}^{c(i+1)+d_{c(i+1)}}|\boldsymbol{s}_i^{i+1}, d_{ci}, d_{c(i+1)}). \nonumber
\end{align}
Note that the final equality in \eref{eq::fano_final} follows from the relative independence of consecutive message blocks and the Markov chain-like behavior of the sequence of drift values. Furthermore, we set $P(s_0=S_0)=1$, since the convolutional code always begins in the same initial state, $S_0$.

Since we only consider drift sequences with initial drift $d_0=0$, we may set $P(d_0=0)=1$ and thus define equivalent decoder metrics for individual branches of the code tree as
\begin{align}
Z(\boldsymbol{v}_0^t \rightarrow \boldsymbol{v}_0^{t+1})&=  \mu(\boldsymbol{v}_0^{t+1})-\mu(\boldsymbol{v}_0^t) \nonumber \\
&=~ -b + \pi_t + \gamma_t(\boldsymbol{y}) + \tau_t(\boldsymbol{y}), \label{eq::fano_br}
\end{align}
where $\tau_t(\boldsymbol{y})$ is a logarithmic ratio of tail probabilities, defined as $\tau_t(\boldsymbol{y})= \log_2 \frac{P_{c(L-t-1)}(\boldsymbol{y}_{c(t+1)+d_{c(t+1)}+1}^N|d_{c(t+1)})}{P_{c(L-t)}(\boldsymbol{y}_{ct+d_{ct}+1}^N|d_{ct})}$. The second equality in \eref{eq::fano_br} follows from $P(s_{t+1}|s_t)=2^{-b}$ and $\pi_i+\gamma_i(\boldsymbol{y})$ can be computed by an iterative process on a lattice structure, outlined in \cite{bahl_decoding_1975, buttigieg_improved_2015}. A similar method can be used to evaluate the logarithmic drift likelihood $\pi_i$ by setting the horizontal, vertical and diagonal edge weights of the lattice to $P_i$, $P_d$ and $P_t$ respectively. Alternatively, one may use the closed-form expression specified in \cite{briffa_timevarying_2014}. The quantity in \eref{eq::pt} can also be evaluated similarly, by making the simplified assumption that all transmitted sequences $\boldsymbol{x} \in \{0,1\}^{R}$ are equally likely. In this case, the horizontal, vertical and diagonal edge weights in the lattice will simply change to $\frac{P_i}{2}$, $P_d$ and $\frac{P_t}{2}$ respectively.

\subsection{Asymptotic Approximation of Branch Metric}

In the upcoming section, we endeavor to derive the computational cutoff rate of Fano's decoder. We do so by establishing an upper bound on the number of computations needed to decode a single information block correctly, in an infinite code tree, i.e., for $L \to \infty$. To simplify the analysis, we use an asymptotic approximation of the tail term $\tau_t(\bfy)$ in the branch metric $Z(\boldsymbol{v}_0^t \rightarrow \boldsymbol{v}_0^{t+1})$, as defined in \eref{eq::fano_br}. 

To proceed along these lines, the methods from \cite{pemantle_analytic_2013, melczer_asymptotics_2020,melczer_invitation_2021} are employed. While the detailed derivation is relegated to Appendix~\ref{app::asymp}, the asymptotic expression of \eref{eq::fano_br} that we finally obtain under the condition $P_i=P_d$, is stated as follows.
\begin{align}
	Z(\boldsymbol{v}_0^t \rightarrow \boldsymbol{v}_0^{t+1})	
	&=  \pi_t+\gamma_t(\boldsymbol{y}) -b+c+d_{c(t+1)}-d_{ct}. \label{eq::br_asym_pipd} 
\end{align}



\begin{rem}
    Equation (\ref{eq::br_asym_pipd}) is quite similar to the expression of Fano branch metric for a binary substitution channel \cite{johannesson_fundamentals_2015}. The same expression is also obtained if we consider a channel that permits only insertions or only deletions, in addition to substitutions.
\end{rem} 

\section{Computational Cut-off Rate} \label{sec::comp_an}
To assess the complexity of Fano's decoder, it suffices to count the number of forward steps taken by the decoder, since each time a new node is visited, branch metrics for all immediate successors must be computed. This is undoubtedly the most costly step in Fano's algorithm. To acquire an upper bound on the total number of visits to nodes in a given code tree, we adopt the approach and the modeling assumptions outlined in \cite{johannesson_fundamentals_2015}, which are summarized in the following.

\subsection{Upper bound on average decoding complexity}

This analysis is conducted under the assumption of correct decoding. Consider a received vector $\boldsymbol{y}_1^N$, which results from the transmission of $L$ blocks, and let the decoding result correspond to the node
$$\boldsymbol{v}_0^{*L}=((s_0^*, d_0^*), (s_1^*, d_1^*), \ldots (s_L^*, d_L^*)).$$

\begin{figure*}[!hbt]
	\centering
	\scalebox{1}{\begin{forest}	
		[ ,name=a,for tree={s sep=18pt,l sep=3cm,dot,grow=0}
		[, name=b1, edge label={\Labeliii}]
		[, name=b2, edge label={\Labeliii}]
		[, name=b3, edge label={\Labeliii}]
		[, name=b4, edge label={\Labeliv}]
		[, name=b5, edge label={\Labeliv}]
		[, name=b6, edge label={\Labeliv}, edge={line width=2pt},
		[, name=c1, edge label={\Labeliii},]
		[, name=c2, edge label={\Labeliii},]
		[, name=c3, edge label={\Labeliii},]
		[, name=c4, edge label={\Labeliv},]
		[, name=c5, edge label={\Labeliv},edge={line width=2pt},
		[, name=d1, edge label={\Labeliii},,edge={line width=2pt}]
		[, name=d2, edge label={\Labeliii},]
		[, name=d3, edge label={\Labeliii},]
		[, name=d4, edge label={\Labeliv},]
		[, name=d5, edge label={\Labeliv},]
		[, name=d6, edge label={\Labeliv}]
		]
		[, name=c6, edge label={\Labeliv}, ]
		]
		]
		\foreach \Nodo in {b1, b2, b3, b4, b5, c1, c2, c3, c4, c6, d1, d2, d3, d4, d5, d6}
		\node[anchor=west,xshift=1.5pt] at (\Nodo) {$\cdots$};
		\node[yshift=-1.3cm,name= tiii,font=\small] at (b1) {$1$};
		\node[name= tii,font=\small] at (a|-tiii) {$0$};
		\node[name= ti,font=\small] at (c1|-tiii) {$2$};
		\node[name= t,font=\small] at (d1|-tiii) {$3$};
		\node[name= tl,font=\small, xshift=-1cm] at (tii) {Depth:};
		\node[xshift=-0.45cm,yshift=-0.05cm,font=\footnotesize] at (a) {$S_0, 0$};
		\node[yshift=-0.25cm,font=\footnotesize] at (b1) {$S_1, 1$};
		\node[yshift=-0.25cm,font=\footnotesize] at (b2) {$S_1, 0$};
		\node[yshift=-0.25cm,font=\footnotesize] at (b3) {$S_1, -1$};
		\node[yshift=0.25cm,font=\footnotesize] at (b4) {$S_0, 1$};
		\node[yshift=0.25cm,font=\footnotesize] at (b5) {$S_0, 0$};
		\node[yshift=0.25cm,xshift=-0.25cm,font=\footnotesize] at (b6) {$S_0, -1$};
		\node[yshift=-0.25cm,font=\footnotesize] at (c1) {$S_1, 0$};
		\node[yshift=-0.25cm,font=\footnotesize] at (c2) {$S_1, -1$};
		\node[yshift=-0.25cm,font=\footnotesize] at (c3) {$S_1, -2$};
		\node[yshift=0.25cm,font=\footnotesize] at (c4) {$S_0, 0$};
		\node[yshift=0.25cm,xshift=-0.25cm,font=\footnotesize] at (c5) {$S_0, -1$};
		\node[yshift=0.25cm,font=\footnotesize] at (c6) {$S_0, -2$};
		\node[yshift=-0.25cm,font=\footnotesize] at (d1) {$S_1, 0$};
		\node[yshift=-0.25cm,font=\footnotesize] at (d2) {$S_1, -1$};
		\node[yshift=-0.25cm,font=\footnotesize] at (d3) {$S_1, -2$};
		\node[yshift=0.25cm,font=\footnotesize] at (d4) {$S_0, 0$};
		\node[yshift=0.25cm,font=\footnotesize] at (d5) {$S_0, -1$};
		\node[yshift=0.25cm,font=\footnotesize] at (d6) {$S_0, -2$};
		\node[name=x1,yshift=0.4cm,xshift=0.8cm,font=\footnotesize] at (a){};
		\node[name=x2,yshift=0.75cm,xshift=-1.0cm,font=\footnotesize] at (x1) {$\substack{\text{Correct path} \\ \text{as per } \boldsymbol{v}_0^{*L}}$ };
		\draw[->] (x2) -- (x1);
		\node[name=s1,yshift=0.6cm,xshift=0.5cm,coordinate] at (d6){};
		\node[name=s2,yshift=-0.5cm,xshift=0.5cm,coordinate] at (d2){};
		\node[name=s3,xshift=-0.55cm,coordinate] at (d2){};
		\node[name=s4,xshift=-0.55cm,coordinate] at (d3){};
		\node[name=s5,xshift=-0.55cm,coordinate] at (d4){};
		\node[name=s6,xshift=-0.55cm,coordinate] at (d5){};
		\node[name=s7,xshift=-0.55cm,coordinate] at (d6){};
		\node[name=s8,xshift=-1.5cm,yshift=0.85cm,font=\footnotesize] at (d6){$\substack{2^\mathrm{nd} \text{ incorrect} \\ \text{subtree, } \tau_2}$};
		\draw[dashed, rounded corners=0.5cm] (s1) -| (s7);
		\draw[dashed] (s7)-- (s6) --(s5)--(s4)--(s3);
		\draw[dashed, rounded corners=0.5cm] (s3) |-(s2);
		\draw[->] (s8) --(s7);
		\node[name=s1,yshift=0.5cm,xshift=0.35cm,coordinate] at (b5){};
		\node[name=s2,yshift=-0.5cm,xshift=0.35cm,coordinate] at (b1){};
		\node[name=s1p,yshift=0.43cm,xshift=0.35cm,coordinate] at (b5){};
		\node[name=s2p,yshift=-0.3cm,xshift=0.35cm,coordinate] at (b4){};
		\node[name=s3p,xshift=-0.45cm,coordinate] at (b5){};
		\node[name=s4p,xshift=-0.45cm,coordinate] at (b4){};
		\node[name=s3,xshift=-0.55cm,coordinate] at (b1){};
		\node[name=s4,xshift=-0.55cm,coordinate] at (b3){};
		\node[name=s5,xshift=-0.55cm,coordinate] at (b4){};
		\node[name=s6,xshift=-0.55cm,coordinate] at (b5){};
		\node[name=s7,xshift=-0.55cm,coordinate] at (b5){};
		\node[name=s8,yshift=-0.5cm,xshift=-1cm,font=\footnotesize] at (s3){$\substack{0^\mathrm{th} \text{ incorrect} \\ \text{subtree, } \tau_0}$};
		\node[name=s5p,yshift=0.27cm,xshift=0.35cm,coordinate] at (b3){};
		\node[name=s5pp,xshift=-0.45cm, yshift=-0.1cm, coordinate] at (b3){};
		\node[name=s1pp,yshift=-0.43cm,xshift=0.35cm,coordinate] at (b1){};
		\node[name=s1ppp,xshift=-0.45cm, coordinate] at (b1){};
		\node[name=s9,yshift=0.3cm,coordinate] at (s3){};
		\draw[dashed, rounded corners=0.5cm] (s1) -| (s7);
		\draw[dashed] (s7)-- (s6) --(s5)--(s4)--(s3);
		\draw[dashed, rounded corners=0.5cm] (s1p) -| (s3p);
		\draw[dashed, rounded corners=0.5cm] (s2p) -| (s4p);
		\draw[dashed] (s3p)-- (s4p);
		\draw[dashed, rounded corners=0.5cm] (s3) |-(s2);
		\draw[->] (s8) -- (s9);
		\draw[dashed, rounded corners=0.5cm] (s1pp) -| (s1ppp);
		\draw[dashed, rounded corners=0.5cm] (s5p) -| (s5pp);
		\draw[dashed] (s1ppp)-- (s5pp);
		\node[name=s8p,yshift=-0.85cm,xshift=-0.2cm, font=\footnotesize] at (s1ppp){$\tau'_0$};
		\node[name=s8pp,xshift=-0.5cm,yshift=0.075cm,coordinate] at (s1pp) {};
		\draw[->] (s8p) -- (s8pp);
		\node[name=s2p, xshift=-1.5cm, yshift=0.5cm, font=\footnotesize] at(s1p){$\tau_0^*$};
		\node[name=s2pp, xshift=-0.75cm, yshift=-0.3cm,coordinate] at(s1p) {};
		\draw[->] (s2p)--(s2pp);
		\node[name=s1,yshift=0.5cm,xshift=0.35cm,coordinate] at (c4){};
		\node[name=s2,yshift=-0.5cm,xshift=0.35cm,coordinate] at (c1){};
		\node[name=s3,xshift=-0.55cm,coordinate] at (c1){};
		\node[name=s4,xshift=-0.55cm,coordinate] at (c2){};
		\node[name=s5,xshift=-0.55cm,coordinate] at (c3){};
		\node[name=s6,xshift=-0.55cm,coordinate] at (c4){};
		\draw[dashed, rounded corners=0.5cm] (s1) -| (s6);
		\draw[dashed] (s6) --(s5)--(s4)--(s3);
		\draw[dashed, rounded corners=0.5cm] (s3) |-(s2);
		\node[name=s1,yshift=0.5cm,xshift=0.35cm,coordinate] at (c6){};
		\node[name=s2,yshift=-0.3cm,xshift=0.35cm,coordinate] at (c6){};
		\node[name=s3,xshift=-0.55cm,coordinate] at (c6){};
		\draw[dashed, rounded corners=0.3cm] (s1) -| (s3);
		\draw[dashed, rounded corners=0.3cm] (s3) |-(s2);
		\node[name=s7,xshift=-1.5cm,yshift=1cm,font=\footnotesize] at (c6){$\substack{1^\mathrm{st} \text{ incorrect} \\ \text{subtree, } \tau_1}$};
		\draw[->] (s7) -- (s3);
		\draw[->] (s7) -- (s6);
		\node[name=s1,xshift=-1.6cm,yshift=-0.65cm,coordinate] at (b6){};
		\node[name=s2,yshift=1cm,xshift=-0.5cm,font=\scriptsize] at (s1){Input};
		\draw[->, dashed] (s2)--(s1);
	\end{forest}}
	{\caption{Incorrect subtrees in a code tree.}\label{fig::ct}}
\end{figure*}
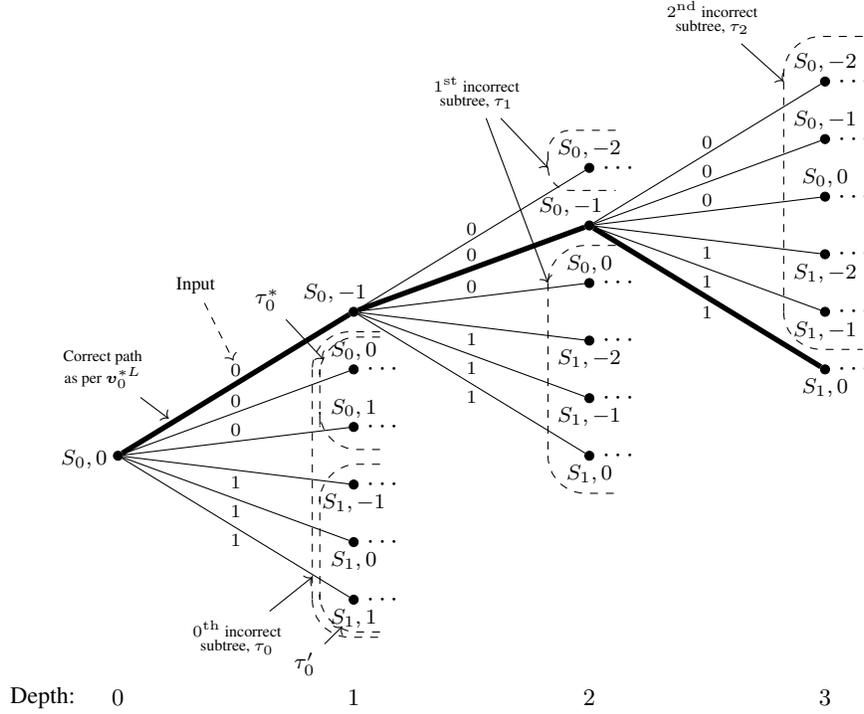

All of the remaining nodes in the joint code and channel tree describe incorrect paths, i.e., paths that predict at least one convolutional code state or drift state inaccurately, and are grouped into $L$ subtrees: $\tau_0, \ldots, \tau_{L-1}$. Here, $\tau_i$ refers to the set of all nodes that hypothesize any false path that stems from $\boldsymbol{v}_0^{i*}$, which is the $i^\mathrm{th}$ node on the correct path as traced by $\boldsymbol{v}_0^{*L}$. We illustrate this partitioning of the joint channel and code tree in Fig. \ref{fig::ct}.

If $C'(\boldsymbol{v}_0^j)$ refers to the number of visits by decoder to node $\boldsymbol{v}_0^j$, then the total complexity of the complete decoding operation can be expressed as
\begin{equation}
	C_{\mathrm{total}}=\sum_{i=0}^{L-1}\big(C'(\boldsymbol{v}^{i*}_0)+\sum_{j=i+1}^{L-1}\sum_{\boldsymbol{v}_0^j \in \tau_i}C'(\boldsymbol{v}_0^j)\big). \nonumber
\end{equation}
We average this quantity over a code ensemble with parameters $(c,b)$, and all possible transmitted and received sequences. For analytical simplicity, we restrict our attention to codes of infinite length. Doing so makes the code tree infinite in length, thereby allowing us to assume similar statistical properties for all $\tau_i$. Under this construct, the task of computational analysis may be reduced to the determination of mean decoding complexity per block, which we define as $C_{\mathrm{av}}={E[C(\boldsymbol{v}^{i*}_0)] = E[C(\boldsymbol{v}^*_0)]}$,
where $C(\boldsymbol{v}^{i*}_0)=C'(\boldsymbol{v}^{i*}_0){+\sum_{j=i+1}^{L-1}\sum_{\boldsymbol{v}_0^j \in \tau_i}C'(\boldsymbol{v}_0^j)}$, i.e., the total number of visits to $\boldsymbol{v}^{i*}_0$ and all nodes in $\tau_0$.

From the earlier discussion on Fano's algorithm, we infer that a drop in metric along $\boldsymbol{v}_0^{*L}$ will cause the decoder to examine alternative paths in the incorrect subtrees. When none of these appear to be promising, the decoder will return to previously visited nodes in $\boldsymbol{v}_0^{*L}$, but with a lowered threshold, and can only move forward when the criterion is upheld. Thus, the number of visits to a node clearly depends on its metric. 

Let $\mu^*_\mathrm{min}$ and $T^*_\mathrm{min}$ denote the minimum node metric and minimum threshold value along $\boldsymbol{v}_0^{*L}$, during a complete decoding operation respectively. It is easy to see that a lower value of $T^*_\mathrm{min}$ implies more backtracking by the decoder to explore incorrect nodes in the tree. Hence, we evaluate $C_{\mathrm{av}}$ as
\begin{align}
	C_\mathrm{av} \! &=\! E[C(\boldsymbol{v}^*_0)] = \! \sum_{i=0}^{\infty} \! P(T^*_\mathrm{min} \! =-i\Delta)E[C(\boldsymbol{v}^*_0)|T^*_\mathrm{min}=-i\Delta]. \label{eq::cav_eq}
\end{align}
The second equality follows from the fact that threshold always starts at $0$, and only changes by the magnitude of $\Delta$. 

With an additional assumption that branch metrics along the correct path are independently and identically distributed, and that the same applies to branch metrics for a given drift change along incorrect paths, we can bound $C_{\mathrm{av}}$ as
\begin{align}
    C_{\mathrm{av}} &\leq C_1 \frac{(1-2^{-\sigma_1 \Delta})2^{-\sigma_0 \Delta}}{1-2^{(\sigma_0+\sigma_1) \Delta}}+C_12^{-\sigma_1 \Delta}+C_3 \frac{2^{-\sigma_0 \Delta}}{1-2^{\sigma_0\Delta}} \nonumber \\
    &+C_2 \frac{(1-2^{-(\sigma_0+\sigma_1)\Delta})2^{-\sigma_0 \Delta}}{1-2^{(\sigma_1+2\sigma_0)\Delta}}+C_2 2^{-(\sigma_0+\sigma_1)\Delta}, \label{eq::cav_bound}
\end{align}
where $C_1$, $C_2$, $C_3$, $\sigma_0$ and $\sigma_1$ are constants depending on the channel and code parameters. The proof, along with more details on the significance of these constants, is relegated to the appendix. The preceding bound only exists when $\sigma_0+\sigma_1<0$ and unfortunately overshoots practical values by several orders of magnitude. However, this bound offers one key insight that $C_\mathrm{av}$ only converges if
\begin{equation}
\sigma_0+\sigma_1<0. \label{eq::ccnew}
\end{equation}

Equations \eref{eq::UB_tau0p} and \eref{eq::tau0_star_final} suggest that $\sigma_1$ is indicative of the rate of rise of computations in $\tau_0$. This leads to an intuitive interpretation of \eref{eq::ccnew}, that the rate of decline of probability $P(\mu^*_\mathrm{min}<y)$, or $\sigma_0$, should sufficiently compensate for the rate of increase of wasteful computations i.e., $\sigma_1$, to allow convergence of average decoding effort per block, i.e., $C_\mathrm{av}$. When \eref{eq::ccnew} holds with equality, the channel essentially operates at computational cutoff rate $R_0$, which is the code rate beyond which using a sequential decoder becomes computationally impractical. This is due to the fact that for rates exceeding $R_0$, the severity of channel noise causes frequent backtracking and more computations in incorrect subtrees. For code trees with infinite depth, $C_\mathrm{av}$ would be unbounded. On the contrary, for code rates below $R_0$, $C_\mathrm{av}$ is bounded, implying that decoding complexity of a single frame grows linearly with tree depth.

\subsection{Error probabilities at cut-off rate}
We now endeavor to briefly describe our method to derive the error probabilities which characterize operation at cutoff rate, given a convolutional code and decoder parameters. As one can surmise from the preceding discussion, this method centers around the equation $\sigma_0+\sigma_1=0$.

In the earlier section, it was stated that $\sigma_0$ and $\sigma_1$ depend on parameters of the channel and the code\footnote{In the following equations, we also note a dependence on decoder parameters $i_{\mathrm{max}}$ and $d_{\mathrm{max}}$.}. More specifically, they satisfy $g_0(\sigma_0)=1$ and $g_1(\sigma_1)=2^{-b}(i_{\mathrm{max}}+d_{\mathrm{max}}+1)^{-1}$, where $i_{\mathrm{max}}$ and $d_{\mathrm{max}}$ are the maximum allowable insertions and deletions per branch, while $g_0(\sigma)$ and $g_1(\sigma)$ are the moment generating functions of the branch metric along correct and incorrect paths, respectively (see Appendix~\ref{app::comp}). Since the current objective is to determine which error probabilities $(P_i,P_d,P_s)$ cause the decoder to operate at cut-off rate, given a specific convolutional code $[c,b,m]$, we deem it appropriate to write $\sigma_0^{(P_i,P_d,P_s)}$ and $\sigma_1^{(P_i,P_d,P_s)}$ instead, so as to explicitly highlight the dependence. To ease the process of finding $\sigma_0^{(P_i,P_d,P_s)}$, the moment generating functions $g_0(\sigma)$ is rearranged as
\begin{align}
g_0(\sigma)&= E[2^{\sigma \zeta^*}] = \sum_{\delta=-d_\mathrm{max}}^{i_\mathrm{max}} \sum_{\substack{\bfx \in \{0,1\}^c \\ \bfy\in \{0,1\}^{c+\delta}}} P(\bfx,\bfy,\delta) 2^{\sigma Z(y|x)} \nonumber \\
&= 2^{-c}\sum_{\delta=-d_\mathrm{max}}^{i_\mathrm{max}} \sum_{\substack{\boldsymbol{x} \in \{0,1\}^c \\ \boldsymbol{y}\in \{0,1\}^{c+\delta}}} P(\boldsymbol{y},\delta|\boldsymbol{x}) 2^{\sigma Z(y|x)}, \label{eq::g0:pxy}
\end{align}
where $Z(\boldsymbol{y}|\boldsymbol{x})$ denotes the asymptotic decoder metric of a branch for the received bits $\boldsymbol{y}$ and transmitted bits $\boldsymbol{x}$. Additionally, the final equality follows from our assumption that all binary vectors of length $c$ are equally likely to be transmitted. Evidently $P(\boldsymbol{y},\delta|\boldsymbol{x})$ is computable by means of the lattice metric, mentioned previously in Section~\ref{sec:decoder:metric}. In a similar manner, we also reformulate $g_1(\sigma)$ to simplify the evaluation of $\sigma_1^{(P_i,P_d,P_s)}$.
\begin{align}
	g_1(\sigma)&=E[2^{\sigma \zeta'}] =\lambda^{-1}\sum_{\delta=-d_{\mathrm{max}}}^{i_{\mathrm{max}}} \sum_{\substack{\boldsymbol{x} \in \{0,1\}^c \\ \boldsymbol{y} \in \{0,1\}^{c+\delta}}} P(\boldsymbol{x}, \boldsymbol{y}) 2^{\sigma Z(y|x)} \nonumber \\
	&=\lambda^{-1}2^{-2c}\sum_{\delta=-d_{\mathrm{max}}}^{i_{\mathrm{max}}} 2^{-\delta} \sum_{\substack{\boldsymbol{x} \in \{0,1\}^c \\ \boldsymbol{y} \in \{0,1\}^{c+\delta}}}  2^{\sigma Z(y|x)}. \label{eq::g1:pxy}
\end{align}

Here, the final equality stems from the fact that along an incorrect path, the received bits are practically independent of the transmitted bits. Furthermore, we assume that all binary vectors of length $c$ are equally likely to have been transmitted, and that all length ($c+\delta$) binary vectors are equally likely to have been received.


\begin{algorithm}[!ht]
	\label{alg::cr}
	\DontPrintSemicolon
	
	\KwInput{Convolutional code parameters $[c,b]$, maximum allowable insertions per branch $i_{\mathrm{max}}$, maximum allowable deletions per branch $d_{\mathrm{max}}$, substitution probability $P_s$, $\alpha$ and $\beta$ such that $P_d=\alpha P_i$ and $P_s=\beta P_i$, error tolerance $\epsilon$.}
	\KwOutput{$p=P_i=\frac{P_d}{\alpha}=\frac{P_s}{\beta}$}
	\Init{}{$b'=b+\log_2 (i_{\mathrm{max}}+d_{\mathrm{max}}+1)$.
		
		Choose $p_a$, $p_b$ such that $p_a<p_b$ and $(\sigma_0^{(p_a,\alpha p_a, \beta p_a)} +\sigma_1^{(p_a,\alpha p_a, \beta p_a)})(\sigma_0^{(p_b,\alpha p_b, \beta p_b)}+\sigma_1^{(p_b,\alpha p_b, \beta p_b)})<0.$

	}
%
	\Do{$|\sigma_0^{(p,\alpha p, \beta p)} +\sigma_1^{(p,\alpha p, \beta p)}|>\epsilon$}
	{
		$p=(p_a+p_b)/2.$
		
		Solve for $\sigma_0^{(p,\alpha p, \beta p)}$ such that $g_0(\sigma)=1$.
		
		Solve for $\sigma_1^{(p,\alpha p, \beta p)}$ such that $g_1(\sigma)=2^{-b'}$.
		
	\If{$\sigma_0^{(p,\alpha p, \beta p)}+\sigma_1^{(p,\alpha p, \beta p)}<0$}
	{
		$p_a=p$.
	}
	\ElseIf{$\sigma_0^{(p,\alpha p, \beta p)} +\sigma_1^{(p,\alpha p, \beta p)}>0$}
	{
		$p_b=p$.
	}
	\Else
	{
	    Stop.
	}
	}
	\caption{Computing error probabilities at cut-off rate}
\end{algorithm}

\emph{Remarks:} From \eref{eq::br_asym_pipd}, \eref{eq::g0:pxy} and \eref{eq::g1:pxy}, we observe that when $P_i=P_d$, it holds that $g_1(\sigma+1)=2^{-b} g_0(\sigma)({i_{\mathrm{max}}+d_{\mathrm{max}}+1})^{-1}$. As a consequence of ${g_0(\sigma_0^{(P_i, P_d, P_s)})=1}$ and ${g_1(\sigma_1^{(P_i, P_d, P_s)}) =2^{-b}(i_{\mathrm{max}}+d_{\mathrm{max}}+1)^{-1}}$, we deduce that $\sigma_1^{(P_i, P_d, P_s)}=\sigma_0^{(P_i, P_d, P_s)}+1$. Consequently, at critical rate, it holds that $\sigma_0=-\frac{1}{2}$ and $\sigma_1=\frac{1}{2}$. Furthermore, this implies that for a channel that guarantees $P_i=P_d$, i.e., $\alpha=1$, it suffices to verify whether a certain value of $p$ upholds $g_0(-\frac{1}{2})=1$, during each iteration of Algorithm \ref{alg::cr}. The same applies to channels where either $P_i=0$, or $P_d=0$.

\begin{figure}[!bht]
	\centering
	\begin{forest}
		[ ,name=a,for tree={s sep=12pt,l sep=6cm,dot,grow'=0}
		[,name=b1dd, edge label={node[above,midway, yshift=1pt,font=\scriptsize]{}}]
		[, name=b1d, edge label={node[above,midway, yshift=1pt,font=\scriptsize]{}}]
		[, name=b1, edge label={node[above,midway, yshift=1pt,font=\scriptsize]{}}]
		[, name=b1i, edge label={node[above,midway,yshift=0.5pt,font=\scriptsize]{}}]
		[, name=b1ii, edge label={node[above,midway,yshift=0.5pt,font=\scriptsize]{}}]
		[, name=b2dd, edge label={node[above,midway,yshift=0.5pt,font=\scriptsize]{}}]
		[, name=b2d, edge label={node[above,midway,yshift=0.5pt,font=\scriptsize]{}}]
		[, name=b2, edge label={node[above,midway,yshift=0.5pt,font=\scriptsize]{}}]
		[, name=b2i, edge label={node[above,midway,yshift=0.5pt,font=\scriptsize]{}}]
		[, name=b2ii, edge label={node[above,midway,yshift=0.5pt,font=\scriptsize]{}}]
		[, name=b3dd, edge label={node[below,midway,yshift=0.5pt,font=\scriptsize]{}}]
		[, name=b3d, edge label={node[below,midway,yshift=0.5pt,font=\scriptsize]{}}]
		[, name=b3, edge label={node[below,midway,yshift=0.5pt,font=\scriptsize]{}}]
		[, name=b3i, edge label={node[below,midway,yshift=1pt,font=\scriptsize]{}}]
		[, name=b3ii, edge label={node[below,midway,yshift=1pt,font=\scriptsize]{}}]
		[, name=b4dd, edge label={node[below,midway,yshift=-1.5pt,font=\scriptsize]{}}]
		[, name=b4d, edge label={node[below,midway,yshift=-1.5pt,font=\scriptsize]{}}]
		[, name=b4, edge label={node[below,midway,yshift=-1.5pt,font=\scriptsize]{}}]
		[, name=b4i, edge label={node[below,midway,yshift=-1.5pt,font=\scriptsize]{}}]
		[, name=b4ii, edge label={node[below,midway,yshift=-1.5pt,font=\scriptsize]{}}]
		]
		\node[yshift=-0.28cm,xshift=-0.2cm,font=\footnotesize] at (a) {$S_0, 0$};
		\node[xshift=0.55cm,font=\footnotesize] at (b1dd) {$S_0, -2$};
		\node[xshift=0.55cm,font=\footnotesize] at (b1d) {$S_0, -1$};
		\node[xshift=0.55cm,font=\footnotesize] at (b1) {$S_0, 0$};
		\node[xshift=0.55cm,font=\footnotesize] at (b1i) {$S_0, 1$};
		\node[xshift=0.55cm,font=\footnotesize] at (b1ii) {$S_0, 2$};
		\node[xshift=0.55cm,font=\footnotesize] at (b2dd) {$S_1, -2$};
		\node[xshift=0.55cm,font=\footnotesize] at (b2d) {$S_1, -1$};
		\node[xshift=0.55cm,font=\footnotesize] at (b2) {$S_1, 0$};
		\node[xshift=0.55cm,font=\footnotesize] at (b2i) {$S_1, 1$};
		\node[xshift=0.55cm,font=\footnotesize] at (b2ii) {$S_1, 2$};
		\node[xshift=0.55cm,font=\footnotesize] at (b3dd) {$S_0, -2$};
		\node[xshift=0.55cm,font=\footnotesize] at (b3d) {$S_0, -1$};
		\node[xshift=0.55cm,font=\footnotesize] at (b3) {$S_0, 0$};
		\node[xshift=0.55cm,font=\footnotesize] at (b3i) {$S_0, 1$};
		\node[xshift=0.55cm,font=\footnotesize] at (b3ii) {$S_0, 2$};
		\node[xshift=0.55cm,font=\footnotesize] at (b4dd) {$S_1, -2$};
		\node[xshift=0.55cm,font=\footnotesize] at (b4d) {$S_1, -1$};
		\node[xshift=0.55cm,font=\footnotesize] at (b4) {$S_1, 0$};
		\node[xshift=0.55cm,font=\footnotesize] at (b4i) {$S_1, 1$};
		\node[xshift=0.55cm,font=\footnotesize] at (b4ii) {$S_1, 2$};
		\node (el1) at ($(a)!0.5!(b1)$) [circle,draw=black,dashed, fill=white, minimum size=0.8cm] {$\substack{000\;000\\(00)}$};
		\node (el2) at ($(a)!0.5!(b2)$)[circle,draw=black,dashed, fill=white, minimum size=0.8cm] {$\substack{000\;111\\(01)}$};
		\node (el3) at ($(a)!0.5!(b3)$)[circle,draw=black,dashed, fill=white, minimum size=0.8cm] {$\substack{111\;011\\(10)}$};
		\node (el4) at ($(a)!0.5!(b4)$)[circle,draw=black,dashed, fill=white, minimum size=0.8cm] {$\substack{111\;100\\(11)}$};
		\node[name=elt1, yshift=3.9cm, xshift=0.5cm, font=\footnotesize] at (a){Encoder output};
		\node[name=elt2, yshift=2cm, xshift=0.5cm, font=\footnotesize] at (a){Input};
		\node [name=elt3, xshift=-5pt, yshift=-2pt] at ($(a)!0.5!(b1d)$){};
		\node [name=elt4, xshift=-1pt, yshift=-8pt] at (elt3){};
		\draw[->] (elt1)--(elt3);
		\draw[->] (elt2)--(elt4);
	\end{forest}
	\caption{Code tree of a $[6,2]$ convolutional code}
	\label{fig::hmm_code2}
\end{figure}
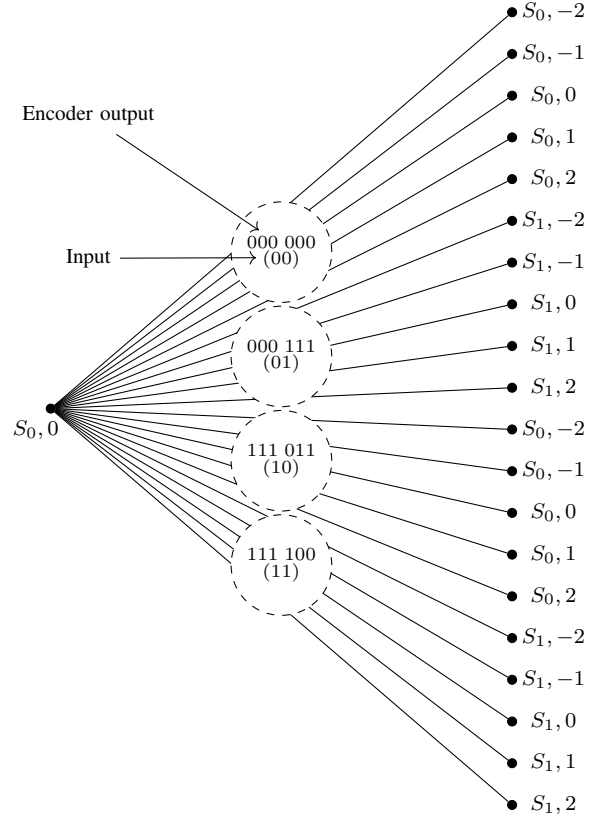

It is also worth noting that for any $j \in \mathbb{N}$, the convolutional codes with parameters $[c,b]$ and $[j c, j b]$ do not induce cut-off rate operation under the same error probabilities $P_i$, $P_d$ and $P_s$, despite possessing the same code rate. In particular, the convolutional code $[j c, j b]$ appears to correspond to a cut-off rate operation under noisier conditions than that for the code $[c,b]$. This is attributable to the path merging phenomenon, which essentially involves the unification of multiple paths when one transforms a joint channel and code tree for a $[c,b]$ code to one for its equivalent $[j c, j b]$. For instance, we can transform the tree in Fig. \ref{fig::hmm_code} into another that corresponds to an equivalent $[6,2,1]$ code, as depicted in Fig. \ref{fig::hmm_code2}. We note that both of these codes will output the same codeword, given the same input sequence.
We can demonstrate the path merging phenomenon by considering any path in Fig.~\ref{fig::hmm_code2}, say $(S_0, 0) \xrightarrow[(01)]{000\;111} (S_1, 0)$. This path actually combines the following paths in the tree shown in Fig.~\ref{fig::hmm_code}.
\begin{alignat*}{3}
	(S_0, 0) &\xrightarrow[(0)]{\quad 000 \quad } &(S_0, -1) \xrightarrow[(1)]{\quad 111 \quad} &(S_1, 0) \nonumber \\
	(S_0, 0) &\xrightarrow[(0)]{\quad 000 \quad} &(S_0, 0) \xrightarrow[(1)]{\quad 111 \quad} &(S_1, 0) \nonumber \\
	(S_0, 0) &\xrightarrow[(0)]{\quad 000 \quad} &(S_0, 1) \xrightarrow[(1)]{\quad 111 \quad} &(S_1, 0) \nonumber
\end{alignat*}
Additionally, it also accounts for the following possibilities
\begin{alignat*}{3}
    (S_0, 0) &\xrightarrow[(0)]{\quad 000 \quad} &(S_0, -3) \xrightarrow[(1)]{\quad 111 \quad} &(S_1, 0) \nonumber \\
	(S_0, 0) &\xrightarrow[(0)]{\quad 000 \quad} &(S_0, -2) \xrightarrow[(1)]{\quad 111 \quad} &(S_1, 0) \nonumber \\
	(S_0, 0) &\xrightarrow[(0)]{\quad 000 \quad} &(S_0, 2) \xrightarrow[(1)]{\quad 111 \quad} &(S_1, 0) \nonumber \\
	(S_0, 0) &\xrightarrow[(0)]{\quad 000 \quad} &(S_0, 3) \xrightarrow[(1)]{\quad 111 \quad} &(S_1, 0) \nonumber 
\end{alignat*}
by virtue of the lattice computation of branch metrics.
\section{Decoding Multiple Sequences} \label{sec::multiple}

When the receiver is given $M$ received sequences, say $\boldsymbol{y}_1, \ldots, \boldsymbol{y}_M$, of lengths $N_1, \ldots, N_M$ respectively, which result from the repeated transmission of the same input sequence over a given channel, we can adapt the sequential decoder to facilitate the simultaneous decoding of these $M$ sequences. This is accomplished by appropriately altering the joint code and channel tree, and the decoder metric. Modifying the former is rather straightforward, in that the HMM that is presumed to produce the sequences $\boldsymbol{y}_1, \ldots, \boldsymbol{y}_M$, now has hidden states, each of which combines the encoder state and $M$ drift variables, one for each of the $M$ received sequences. More explicitly, we let $d_{i,j}$ indicate the drift of the $j^{\mathrm{th}}$ received sequence after the transmission of $i$ bits.

\subsection{Decoder metric}
 
Akin to its initial definition in (\ref{eq::fanoL}), the decoder metric is now modified to account for multiple sequences, as follows.
\begin{align}
	\mu(\boldsymbol{v}_0^t)
	&=\log_2 P(\boldsymbol{v}_0^t, \boldsymbol{y}_1, \ldots, \boldsymbol{y}_M) -\log_2 P(\boldsymbol{y}_1, \ldots, \boldsymbol{y}_M), \label{eq::fano_mult}
\end{align}
where $\boldsymbol{v}_0^t$ represents a node in the joint code and channel tree at depth $t$, that is reached from the root via the sequence of encoder states $(s_0, \ldots, s_t)$, and the sequence of drift state vectors $(\bfd_0, \bfd_{c}, \ldots, \bfd_{ct})$, wherein the vector $\bfd_{ci}=(d_{ci,1}, \ldots, d_{ci,M})$ specifies the drift of each of the $M$ received sequence after the transmission of $ci$ bits. As before, the initial drifts $d_{0,1}, \ldots, d_{0,M}$ are set to zero, and we choose the following vector representation for $\boldsymbol{v}_0^t$.
\begin{align*}
    \bfv_0^t &=((s_0,d_{0,1}, \ldots, d_{0,M}), (s_1,d_{c,1},\ldots, d_{c,M}), \\
    & \qquad \ldots, (s_t,d_{ct,1}, \ldots, d_{ct,M})).
\end{align*}

As done before, we recognize that the path traced from the root by the node $\boldsymbol{v}_0^t$ does not account for the entire transmitted sequence, but only its first $t$ blocks, which correspond to the partial received sequences $(\boldsymbol{y}_1)_1^{ct+d_{ct,1}}, \ldots, (\boldsymbol{y}_M)_1^{ct+d_{ct,M}}$. The remainder of these received sequences is assumed to have been produced by a tailing message sequence that causes the encoder to traverse the code states $\tilde{\boldsymbol{s}}_{t+1}^L=(\tilde{s}_{t+1}, \ldots, \tilde{s}_{L})$. With the additional assumption that the remaining received bits are independent of the previously transmitted bits, we proceed to expand the terms in (\ref{eq::fano_mult}).
\begin{align}
	&P(\boldsymbol{v}_0^t, \boldsymbol{y}_1, \ldots, \boldsymbol{y}_M) =\sum_{ \tilde{\boldsymbol{s}}_{t+1}^L } P(\boldsymbol{v}_0^t, \tilde{\boldsymbol{s}}_{t+1}^L, \boldsymbol{y}_1, \ldots, \boldsymbol{y}_M) \nonumber \\
	&=\sum_{\tilde{\boldsymbol{s}}_{t+1}^L } P(\boldsymbol{v}_0^t)P(\tilde{\boldsymbol{s}}_{t+1}^L) P(\boldsymbol{y}_1, \ldots, \boldsymbol{y}_M | \boldsymbol{v}_0^t,\tilde{\boldsymbol{s}}_{t+1}^L) \nonumber \\
	&=\sum_{\tilde{\boldsymbol{s}}_{t+1}^L } P(\boldsymbol{v}_0^t)P(\tilde{\boldsymbol{s}}_{t+1}^L) \prod_{i=1}^{M} P(\boldsymbol{y}_i| \boldsymbol{v}_0^t,\tilde{\boldsymbol{s}}_{t+1}^L) \label{eq::fano_m_yind} \\
	&= P(\boldsymbol{v}_0^t) \! \prod_{i=1}^{M}\!  P((\boldsymbol{y}_i)_1^{ct+d_{ct,i}}| \boldsymbol{v}_0^t)\! \sum_{\tilde{\boldsymbol{s}}_{t+1}^L }\!\! P\Big((\boldsymbol{y}_i)_{ct+d_{ct,i}+1}^{N_i}, \tilde{\boldsymbol{s}}_{t+1}^L| d_{ct,i}\Big) \nonumber \\
	&= P(\boldsymbol{v}_0^t) \prod_{i=1}^{M}  P\Big((\boldsymbol{y}_i)_1^{ct+d_{ct,i}}| \boldsymbol{v}_0^t\Big)  P_{c(L-t)}\Big((\boldsymbol{y}_i)_{ct+d_{ct,i}+1}^{N_i}| d_{ct,i}\Big), \nonumber
\end{align}
where (\ref{eq::fano_m_yind}) follows from the independence of the $M$ received sequences of each other, given a complete path in the joint code and channel tree. We can also simplify the joint probability of the $M$ received sequences by marginalizing it over all possible transmitted sequences that might produce them.
\begin{align}
	P(\boldsymbol{y}_1, \ldots, \boldsymbol{y}_M)&= \sum_{\boldsymbol{x} \in \{0,1\}^{cL}} P(\boldsymbol{x}) P(\boldsymbol{y}_1, \ldots, \boldsymbol{y}_M| \boldsymbol{x}) \nonumber \\
	&= \prod_{i=1}^{M} \sum_{\boldsymbol{x} \in \{0,1\}^{cL}} P(\boldsymbol{x}) P(\boldsymbol{y}_i| \boldsymbol{x}) = \prod_{i=1}^{M} P_{cL} (\boldsymbol{y}_i). \nonumber
\end{align}

Thus, the decoder metric of node $\bfv_0^t$ can be computed as
\begin{align}
\mu(\boldsymbol{v}_0^t)&= \log_2 P(\boldsymbol{v}_0^t) + \sum_{i=1}^{M} \bigg(\log_2 P\Big((\boldsymbol{y}_i)_1^{ct+d_{ct,i}}| \boldsymbol{v}_0^t\Big) \nonumber \\
&+ \log_2 P_{c(L-t)}\Big((\boldsymbol{y}_i)_{ct+d_{ct,i}+1}^{N_i} | d_{ct,i}\Big)- \log_2  P_{cL} (\boldsymbol{y}_i) \bigg). \nonumber
\end{align}

As in (\ref{eq::fano_br}), we can also derive the branch decoder metric for multiple sequences.
\begin{align}
	Z(\boldsymbol{v}_0^{t} \rightarrow &\boldsymbol{v}_0^{t+1})=  \mu(\boldsymbol{v}_0^{t+1})-\mu(\boldsymbol{v}_0^{t}) \nonumber \\
	&= -b + \sum_{i=1}^{M} \bigg( \pi^{(i)}_t  +\gamma^{(i)}_t(\boldsymbol{y}_i)+ \tau^{(i)}_t(\boldsymbol{y}_i) \bigg), \label{eq::fano_br_mult}
\end{align}
where, analogous to \eref{eq::fano_br}, we define $\pi^{(i)}_t$, $\gamma^{(i)}_t(\boldsymbol{y}_i)$ and $\tau^{(i)}_t(\boldsymbol{y}_i)$ as follows.
\begin{align*}
    \pi^{(i)}_t&=\log_2 \! P(d_{c(t+1),i}|d_{ct,i}) \\
    \gamma^{(i)}_t(\boldsymbol{y}_i)&=\log_2 P\Big( (\boldsymbol{y}_i)_{ct+d_{ct,i}+1}^{c(t+1)+d_{c(t+1),i}}| \boldsymbol{s}_t^{t+1}, d_{ct,i}, d_{c(t+1),i}\Big) \\
    \tau^{(i)}_t(\boldsymbol{y}_i)&= \log_2 \frac{P_{c(L-t-1)}\big((\boldsymbol{y}_i)_{c(t+1)+d_{c(t+1),i}+1}^{N_i}\Big)}{P_{c(L-t)}\Big((\boldsymbol{y}_i)_{ct+d_{ct,i}+1}^{N_i}\Big)}.
\end{align*}
\subsection{Computational analysis}

The process of evaluating the complexity of sequential decoding in case of multiple sequences is exactly the same as described in Section~\ref{sec::comp_an}, except that the expression for the branch decoder metric is now given by (\ref{eq::fano_br_mult}). This expression is actually related to the branch decoder metric for a single sequence as follows.
\begin{align}
	&Z(\boldsymbol{v}_0^{t} \rightarrow \boldsymbol{v}_0^{t+1})\! = b(M-1)\! + \! \sum_{i=1}^{M} \! \bigg(\hspace{-2mm} -b+ \pi^{(i)}_t  + \gamma^{(i)}_t  +\tau^{(i)}_t(\boldsymbol{y}_i) \hspace{-1.2mm}\bigg) \nonumber \\
	&=b(M-1)+\sum_{i=1}^{M} Z\Big((\boldsymbol{v}_0^{t})_i \rightarrow (\boldsymbol{v}_0^{t+1})_i\Big), \label{eq::fan_br_msplit}
\end{align}
where $(\boldsymbol{v}_0^{t})_i$ represents a path in the joint code and channel tree for decoding only $\boldsymbol{y}_i$, that starts from the root and ends in a node at depth $t$, while traversing the encoder states $(s_0, \ldots, s_t)$, i.e., same as $\boldsymbol{v}_0^t$, and drift states for the $i^{th}$ received sequence in $\boldsymbol{v}_0^t$, i.e., $(d_{0,i}, d_{c,i}, \ldots, d_{ct,i})$. Naturally, $Z\Big((\boldsymbol{v}_0^{t})_i \rightarrow (\boldsymbol{v}_0^{t+1})_i\Big)$ is simply the decoder metric of the branch $(\boldsymbol{v}_0^{t})_i \rightarrow (\boldsymbol{v}_0^{t+1})_i$, if one was performing sequential decoding of the sequence $\boldsymbol{y}_i$ alone.

By exploiting (\ref{eq::fan_br_msplit}), the moment generating functions $g_0(\sigma)$ and $g_1(\sigma)$ can be suitably extended for the case of multiple sequences, i.e., $g_0^{(M)}(\sigma)=2^{\sigma b (M-1)} (g_0(\sigma))^M$ and $g_1^{(M)}(\sigma)=2^{\sigma b (M-1)} (g_1(\sigma))^M$.

For a given convolutional code, cutoff rate operation will occur for insertion, deletion and substitution probabilities if it holds that $\sigma_0^{(P_i, P_d, P_s)}+\sigma_1^{(P_i, P_d, P_s)}=0$, where 
$g^{(M)}_0(\sigma_0^{(P_i, P_d, P_s)})=1$ and $g^{(M)}_1(\sigma_1^{(P_i, P_d, P_s)}) {=2^{-b-\log_2(i_{\mathrm{max}}+d_{\mathrm{max}}+1)}}$. This implies that in order to determine the values of $P_i$, $P_d$ and $P_s$ that induce cut-off rate operation during decoding of multiple sequences for a given convolutional code, we can simply utilize Algorithm \ref{alg::cr} following minor adjustments.
\section{Results} \label{sec::results}
To evaluate this decoding strategy, we use three standard, rate $1/3$ convolutional codes with distinct constraint lengths, as outlined in Table \ref{tab::codes}. For $L=300$ information blocks, rates of terminated codewords of CC1, CC2 and CC3 are $0.332$, $0.327$ and $0.323$ respectively.
\begin{figure}[t]%
	\centering%
	\begin{tikzpicture}
	\begin{axis}[legend style={nodes={scale=0.9, transform shape}},
	xmode=log, ymode=log,
	xmin=2e-3, xmax=1,
	ymin=1e-8, ymax=0.07,
	xlabel=\textsc{$P_i=P_d$}, ylabel={BER},
	legend pos=south east, legend style={font=\footnotesize},
	grid=both]
	\addplot plot [blue, solid, mark=square*,mark options={fill=blue}]coordinates {          
		(3e-3, 0.000458)
		(5e-3, 0.000848)  
		(1e-2, 0.00246333)
		(2e-2, 0.008132)
		(4e-2, 0.03044561)			
	};
	\addlegendentry{CC1 Fano $P_s=0$}
	\addplot plot[blue, dashed, mark=square*, mark options={fill=blue}] coordinates {          
		(3e-3, 0.00035667)
		(5e-3, 0.000642)  
		(1e-2, 0.00171133)
		(2e-2, 0.00546733)
		(4e-2, 0.021534)	
	};
	\addlegendentry{CC1 Viterbi $P_s=0$}
	\addplot plot[red, solid, mark=triangle*, mark options={fill=red}] coordinates {          
		(3e-3, 0.0000280)  
		(5e-3, 0.000074)
		(1e-2, 0.00027600)
		(2e-2, 0.00160790)
	};
	\addlegendentry{CC2 Fano $P_s=0.02$}
	\addplot plot[red, dashed, mark=triangle*, mark options={fill=red}] coordinates {
		(3e-3, 0.000006670)          
		(5e-3, 0.00001667)  
		(1e-2, 0.00007533)
		(2e-2, 0.00075933)
		(4e-2, 0.01070133)
	};
	\addlegendentry{CC2 Viterbi $P_s=0.02$}
	\addplot plot[red, solid, mark=square*, mark options={fill=red}] coordinates { 
		(5e-3, 0.0000006667)      
		(1e-2, 0.0000055556)
		(2e-2, 0.00008200)
		(4e-2, 0.00185989)
	};
	\addlegendentry{CC2 Fano $P_s=0$}	
	\addplot plot[red, dashed, mark=square*, mark options={fill=red}] coordinates {
		(1e-2, 2e-6)          
		(2e-2, 0.00002267)
		(4e-2, 0.001278)  
	};
	\addlegendentry{CC2 Viterbi $P_s=0$}
    \addplot plot[red, dash dot, mark=square*, mark options={fill=red}] coordinates
	{
        (2e-2, 1.5416e-5)
	    (4e-2, 0.00093458)
        (5e-2, 0.00364733)
		(6e-2, 0.0118)
	};
	\addlegendentry{CC2 BCJR $P_s=0$ [13]}
	\addplot plot[brown, solid, mark=square*, mark options={fill=brown}] coordinates {          
		(2e-2,  0.000002)
		(3e-2, 0.0000337778)
		(4e-2, 0.00013765)
	};
	\addlegendentry{CC3 Fano $P_s=0$ }
	\end{axis}
	\end{tikzpicture}
	\caption{Comparison of bit error rates of $[3,1]$ terminated codes consisting of $L=300$ information blocks, with varying memory.}
	\label{fig::ber_memory}
	\vspace{-2mm}
\end{figure}
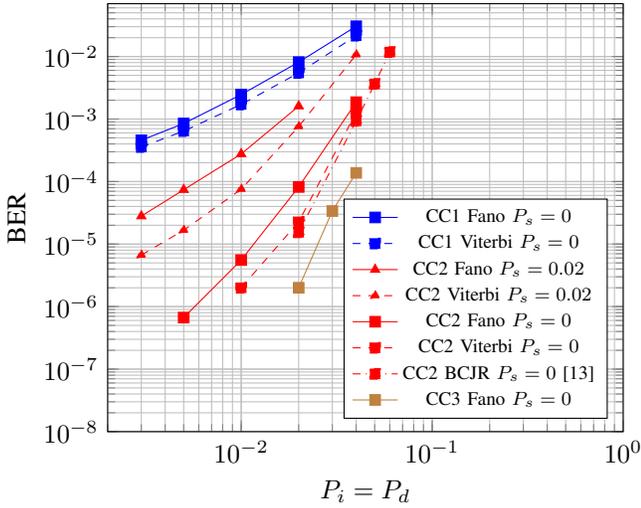%

\begin{figure}[t]
    \centering
     
     \begin{tikzpicture}
	\begin{axis}[legend style={nodes={scale=0.9, transform shape}},
	xmode=log, ymode=log,
	xmin=2e-3, xmax=0.5,
	ymin=2e-7, ymax=0.2,
	xlabel=\textsc{$P_i=P_d$}, ylabel={BER},
	legend pos=south east, legend style={font=\footnotesize},
	grid=both]
	\addplot plot[blue, dashed, mark=square*, mark options={fill=blue}] coordinates {
	    (3e-3, 0.000458)
		(5e-3, 0.000848)  
		(1e-2, 0.00246333)
		(2e-2, 0.008132)
		(4e-2, 0.03044561)
	};
	\addlegendentry{CC1 $M=1$}
	\addplot plot[blue, solid, mark=square*, mark options={fill=blue}] coordinates {          
		(3e-3, 0.0000416667)
		(5e-3, 0.0000986667)  
		(1e-2, 0.0003436667)
		(2e-2, 0.0011630000)

	};
	\addlegendentry{CC1 $M=2$}
	\addplot plot[blue, solid, mark=triangle*, mark options={fill=blue}] coordinates {          
		(3e-3, 0.00000333)
		(5e-3, 0.00001533)  
		(1e-2, 0.00008067)
		(2e-2, 0.00042000)
		(3e-2, 0.00106267)
	};
	\addlegendentry{CC1 $M=3$}
	\addplot plot[red, dashed, mark=square*, mark options={fill=red}] coordinates {
		(5e-3, 0.0000006667)      
		(1e-2, 0.0000055556)
		(2e-2, 0.00008200)
		(4e-2, 0.00185989)
	};
	\addlegendentry{CC2 $M=1$}
	\addplot plot[red, solid, mark=square*, mark options={fill=red}] coordinates { 
		(2e-2, 0.0000056667)
		(3e-2,  0.00002833) 
		(4e-2, 0.0000756667) 
		(0.06, 0.00031247) 
	};
	\addlegendentry{CC2 $M=2$}
	\addplot plot[red, solid, mark=triangle*, mark options={fill=red}] coordinates
	{ 
		(2e-2, 0.00000150)
		(2.5e-2, 0.00000500)
		(3e-2, 0.0000093333)
		(3.5e-2, 0.00001853)
	};
	\addlegendentry{CC2 $M=3$}
	\end{axis}
	\end{tikzpicture}
    	\caption{Comparison of bit error rates of $[3,1]$ terminated codes consisting of $L=300$ information blocks, using Fano's decoder given single and multiple sequences. $P_s=0$.}
    \label{fig:ber_multiple}
\end{figure}
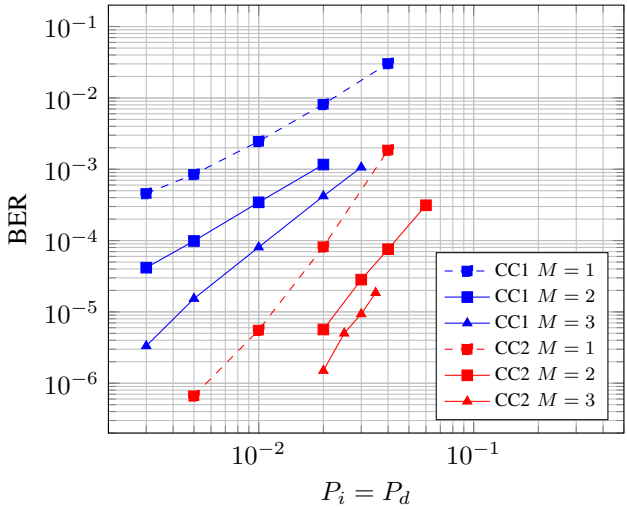

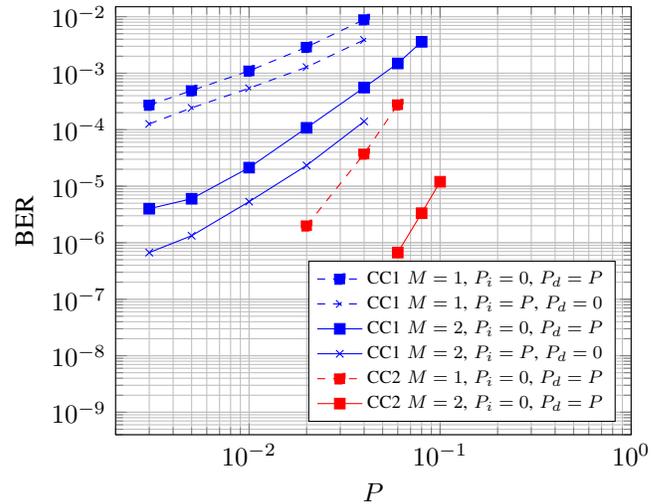
\begin{figure}[t]
    \centering
     
     \begin{tikzpicture}
	\begin{axis}[legend style={nodes={scale=0.86, transform shape}},
	xmode=log, ymode=log,
	xmin=2e-3, xmax=1,
	ymin=4e-10, ymax=0.015,
	xlabel=\textsc{$P$}, ylabel={BER},
	legend pos=south east, legend style={font=\footnotesize},
	grid=both]
 \addplot plot[blue,dashed, mark=square*, mark options={fill=blue}] coordinates {
		(3e-3, 0.00027333)
		(5e-3, 0.00049267)  
		(1e-2, 0.00110067)
		(2e-2, 0.00289133)
		(4e-2, 0.00888000)
	};
	\addlegendentry{CC1 $M=1$, $P_i=0$, $P_d=P$}
	\addplot plot[blue,dashed, mark=x, mark options={fill=blue}] coordinates {
		(3e-3, 0.00012667)
		(5e-3, 0.00024333) 
		(1e-2, 0.000542)
		(2e-2, 0.00128133)
		(4e-2, 0.00388733)
	};
	\addlegendentry{CC1 $M=1$, $P_i=P$, $P_d=0$}
 \addplot plot[blue, solid, mark=square*, mark options={fill=blue}] coordinates {
		(3e-3, 0.00000400)
		(5e-3, 0.00000600)  
		(1e-2, 0.00002133)
		(2e-2, 0.00010800)
		(4e-2, 0.00055800)
		(6e-2, 0.00148933)
		(8e-2, 0.00359933)
	};
	\addlegendentry{CC1 $M=2$, $P_i=0$, $P_d=P$}
	\addplot plot[blue, solid, mark=x, mark options={fill=blue}] coordinates {
		(3e-3, 0.00000067)
		(5e-3, 0.00000133)  
		(1e-2, 0.00000533)
		(2e-2, 0.00002333)
		(4e-2, 0.00014011)
	};
	\addlegendentry{CC1 $M=2$, $P_i=P$, $P_d=0$}
	\addplot plot[red, dashed, mark=square*, mark options={fill=red}] coordinates {
		(2e-2, 0.00000200)
		(4e-2, 0.00003733)
		(6e-2, 0.00027733)
	};
	\addlegendentry{CC2 $M=1$, $P_i=0$, $P_d=P$}
	\addplot plot[red, solid, mark=square*, mark options={fill=red}] coordinates {
		(6e-2, 0.00000067)
		(8e-2, 0.00000333 )
		(0.1, 0.00001200)
	};
	\addlegendentry{CC2 $M=2$, $P_i=0$, $P_d=P$}
	\end{axis}
	\end{tikzpicture}
    	\caption{Comparison of bit error rates of $[3,1]$ terminated codes consisting of $L=300$ information blocks, using Fano's decoder, given single and multiple sequences, when either $P_i=0$ or $P_d=0$; and $P_s=0$.}
    \label{fig:ber_pid=0}
\end{figure}

\begin{table}[b]
	\setlength{\tabcolsep}{13.2pt}
	\centering
	\caption{Convolutional codes for simulations}
	\renewcommand{\arraystretch}{1.1}
	\begin{tabular}{clcc} \specialrule{1.2pt}{0pt}{0pt}
		Code & $[c,b,m]$ & Gen. polynomial & $d_{\mathrm{free}}$ \\ \specialrule{.8pt}{0pt}{0pt}
		CC1 & $[3,1,1]$ & $1 \quad 3 \quad 3$ & 5\\
		CC2 & $[3,1,6]$ & $117 \quad 127 \quad 155$ & 15 \\
		CC3 & $[3,1,10]$ & $3645 \quad 2133 \quad 3347$ & 21 \\
		\specialrule{1.2pt}{0pt}{4pt}
	\end{tabular}\label{tab::codes}
\end{table}
The generator polynomials are stated in octal form. We simulated the transmission of terminated codewords with $L=300$ information blocks and offset by a random sequence\footnote{Helps to maintain synchronization, like marker or watermark codes \cite{buttigieg_improved_2015}.}, over a channel with parameters set to either $P_i=P_d$, $P_i=0$ or $P_d=0$. Additionally, the substitution probability was set to either $P_s=0$ or $P_s=0.02$.  To limit decoder complexity, we constrain the maximum allowable insertions/deletions per block to $c$, and ignore drift states of magnitudes exceeding $30$. Additionally, if the number of forward steps exceeds $10^5$, the decoder is terminated and the number of bit errors in the partial output is counted, while the bits absent from the output are considered to be in error.

\subsection{Decoding performance}
We assess the decoding accuracy of Fano's sequential decoder by measuring bit error rates over a range of $P_i=P_d$ and $P_s=0,0.02$, and subsequently perform a comparison with Viterbi decoder, as depicted in Fig. \ref{fig::ber_memory}\footnote{A direct comparison with \cite{mansour_convolutional_2002} was not possible since we could not extract the data points in its plots with sufficient precision. Furthermore, the trellis employed in \cite{mansour_convolutional_2002} differs markedly from the model used here, in that \cite{mansour_convolutional_2002} only permits one insertion at each stage and allows edges to skip over a time step. Finally, \cite{buttigieg_improved_2015} demonstrated the superiority of the lattice metric used for the decoder metric in (\ref{eq::fano_br}), over the one defined in \cite{mansour_convolutional_2002}.}. Since it only partially examines a given code tree, Fano's decoder is inherently sub-optimal and is thus outperformed by the Viterbi decoder. Unsurprisingly, a higher value of $P_s$ worsens the frame error rate for both decoders.

We also investigate the impact of using $M=2,3$ sequences over a single sequence. As illustrated in Fig. \ref{fig:ber_multiple}, the knowledge of even one additional received sequence can significantly improve decoder performance. Fig. \ref{fig:ber_pid=0} suggests the same, under the conditions either $P_i=0$ or $P_d=0$; and $P_s=0$. Fig. \ref{fig:ber_pid=0} also shows that deletions affect the bit error rate more than insertions, for the same occurrence probability. 

\subsection{Simulated complexity}
The practical complexities of the two decoders are compared in terms of the number of branch metric computations performed. So, for a Viterbi decoder, we compute the total number of branches in the associated trellis, while for Fano's algorithm, it suffices to measure the average number of forward steps taken, denoted by $\mathcal{F}_\mathrm{av}$, and to multiply this with the number of outgoing edges per node, as explained in Section~\ref{sec::comp_an}. For the former case, we limit our attention to the initial non-terminating part of the trellis. Since in either case, a single node produces $2^b(i_\mathrm{max}+d_\mathrm{max}+1)$ outgoing branches, we choose to define the complexity reduction factor as $\nu= \mathcal{N}_\mathrm{tot} / \mathcal{F}_\mathrm{av}$, where $\mathcal{N}_\mathrm{tot}$ denotes the total number of nodes in the trellis.
\begin{figure}[t]%
		\begin{tikzpicture}
			\begin{axis}[legend style={font=\footnotesize, nodes={scale=0.86, transform shape}},
				xmode=log, ymode=log,
				xmin=20e-4, xmax=0.3,
				ymax=2e6, ymin=2,
				xlabel=\textsc{$P_i=P_d$}, ylabel={$\nu$},
				legend pos=north east,
				extra x ticks={0.01922, 0.0057},
				extra x tick labels={$P^*_0$,$P^*_{0.02}$},
				extra x tick style = {
					red,
					font=\bfseries
				},
				grid=both]
				\addplot plot[blue,mark=square*, solid] coordinates {          
					(3e-3, 104.8786)  
					(5e-3, 102.9429)  
					(1e-2, 93.6214)
					(2e-2, 55.8714)
					(4e-2,  6.3786)	
				};
				\addlegendentry{CC1 at $P_s=0$, $M=1$}
				\addplot plot[red,mark=square*,mark options={color=red},solid] coordinates {          
					(3e-3, 3.2923e+03)
					(5e-3, 3.2157e+03)
					(1e-2, 2702)  
					(2e-2, 1241.9)
					(4e-2, 88.3571)
				};
				\addlegendentry{CC2 at $P_s=0$, $M=1$}
				\addplot plot[red, solid, mark=triangle*, mark options={fill=red}] coordinates  {          
					(3e-3, 2223)
					(5e-3, 1.8987e+03)
					(1e-2, 795.7143)  
					(2e-2, 174.4286)
				};
				\addlegendentry{CC2 at $P_s=0.02$, $M=1$}
				\addplot plot[brown, solid, mark=square*, mark options={fill=brown}] coordinates  {          
					(3e-3, 5.3644e4)
					(5e-3, 5.2148e4)
					(1e-2, 4.5109e4)  
					(2e-2, 1.9432e4)
					(3e-2, 5.1843e3)
					(4e-2, 1.6371e3)
				};
				\addlegendentry{CC3 at $P_s=0$, $M=1$}
				
				
				
				\draw [thick, dotted, red] (0.01922,4e7) -- (0.01922,0);
				\draw [thick, dotted, red] (0.0057,4e7) -- (0.0057,0);
			\end{axis}%
	\end{tikzpicture}%
	\caption{Complexity reduction factor of Fano's decoder for $M=1$ received sequence and $[3,1]$ terminated codes consisting of $L=300$ information	blocks.}%
	\label{fig::nu}%
	\vspace{-2mm}
\end{figure}
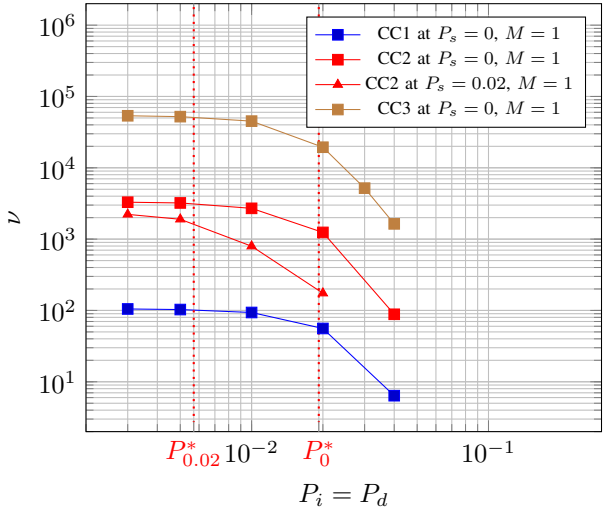%

\begin{figure}[t]%
		\begin{tikzpicture}
			\begin{axis}[legend style={font=\footnotesize, nodes={scale=0.86, transform shape}},
				xmode=log, ymode=log,
				xmin=20e-4, xmax=0.2,
				ymax=1e6, ymin=1e2,
				xlabel=\textsc{$P_i=P_d$}, ylabel={$\nu$},
				legend pos=north east,
				extra x ticks={0.032604},
				extra x tick labels={$P^{2*}_0$},
				extra x tick style = {
					red,
					font=\bfseries
				},
				grid=both]
				\addplot plot[blue,mark=square*, solid] coordinates {          
					(3e-3, 6.8817e3)  
					(5e-3, 6.6193e3)  
					(1e-2, 5.7642e3)
					(2e-2, 3.4289e3)
					(3e-2, 1.7572e3)
					(4e-2, 0.7137e3)
					(5e-2, 0.2952e3)
				};
				\addlegendentry{CC1 at $P_s=0$}
				\addplot plot[red,mark=square*,mark options={color=red},solid] coordinates {          
					(3e-3, 2.1784e5)
					(5e-3, 1.9507e5)
					(1e-2, 1.6432e5)  
					(2e-2, 0.8025e5)
					(3e-2, 0.2789e5)
					(4e-2, 0.1275e5)
					(5e-2, 0.0545e5)
				};
				\addlegendentry{CC2 at $P_s=0$}
				\draw [thick, dotted, red] (0.032604,4e7) -- (0.032604,0);
			\end{axis}%
	\end{tikzpicture}%
	\caption{Complexity reduction factor of Fano's decoder for $M=2$ received sequences and $[3,1]$ terminated codes consisting of $L=300$ information	blocks.}%
	\label{fig::nu_2}%
	\vspace{-2mm}
\end{figure}
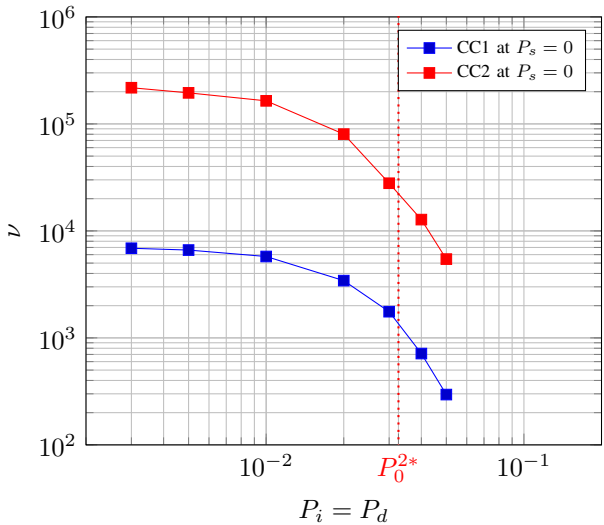%

Fig. \ref{fig::nu} demonstrates significant reductions in decoding effort by the use of Fano's decoder, particularly for low error probabilities. We also notice that as channel noise deteriorates, especially beyond $P_0^*$ and $P_{0.02}^*$ which mark operation at cutoff rate for substitution probabilities $P_s=0$ and $P_s=0.02$ respectively, Fano's decoder gradually loses its computational merit. This is because as error probabilities increase, the decoder is forced to examine many more incorrect paths before the correct one is found. Similarly, Fig. \ref{fig::nu_2} exhibits the amount of computational effort saved by Fano's decoder in the context of decoding $M=2$ received sequences simultaneously. $P^{2*}_0$ marks the value of $P_i=P_d$ at which cut-off rate operation occurs for $P_s=0$, $c=3$ and $b=1$.

\section{Conclusion}

In this work, we extended Fano's sequential decoder for use with channels that allow insertion, deletion as well as substitution errors. To this end, we obtain a new decoding metric that serves as a generalization of the Fano metric, originally introduced in \cite{fano_heuristic_1963}. By deriving an asymptotic approximation of this decoder metric, we also assess the average computational complexity of the decoder. This in turn leads to an algorithm for the determination of channel noise levels which correspond to cut-off rate operation, given a specific set of convolutional code parameters. Furthermore, we repeat the derivation of the decoder metric and computational analysis for the case of multiple sequences. Finally, simulation results reveal that though the proposed decoder performs sub-optimally in comparison to a Viterbi decoder, it offers significant reductions in computational effort, especially for convolutional codes with high memory. 

\printbibliography

\appendix

\subsection{Asymptotic approximation of Tail Probabilities} \label{app::asymp}
It is easy to see from \eref{eq::fano_br} that to obtain the asymptotic expression of the branch decoder metric, it suffices to derive the same $\tau_t(\boldsymbol{y})$, which is a ratio of tail probabilities. As stated earlier, a tail probability $P_R(\boldsymbol{y}_1^N)$ can be seen as a weighted sum of paths through a two-dimensional lattice. These paths start at origin $(0,0)$, and can progress towards the endpoint $(R,N)$ by taking one step at a time, either horizontally, vertically or diagonally, with the respective weights of these edges being $\frac{P_i}{2}$, $P_d$ and $\frac{P_t}{2}$. This definition strongly resembles the concept of a weighted Delannoy number \cite{comtet, why_del, fray, fray2}, except for one key difference in that the computation of $P_R(\boldsymbol{y}_1^N)$ does not involve any horizontal edge in the final row of the lattice. We define a weighted Delannoy number $d_{r,s}$ as the sum of weights of all possible paths in a two-dimensional lattice with only horizontal, vertical and diagonal edges, from the starting point $(0,0)$ to the endpoint $(r,s)$. Referring to the horizontal, vertical and diagonal edge weights as $\alpha$, $\beta$ and $\gamma$ respectively, we note that $d_{r,s}$ can be specified by the recurrence relation
\begin{equation}
    d_{r,s}=\alpha d_{r,s-1} +\beta d_{r-1,s}+  \gamma d_{r-1, s-1}, \label{eq::wdel}
\end{equation}
and is subject to the initial conditions $d_{0,0}=1$, $d_{r,0}=\beta^r$ and $d_{0,s}=\alpha^s$ for all $r\geq 0$ and $s\geq 0$.

As the lattice structure used to compute $P_M(\boldsymbol{y}_1^N)$ does not permit any horizontal edges in the last line, we deduce that
\begin{equation}
    P_R(\boldsymbol{y}_1^N)=d_{R,N}-\frac{P_i}{2} d_{R-1,N}, \label{eq::ptail_del}
\end{equation}
if $\alpha=\frac{P_i}{2}$, $\beta=P_d$ and $\gamma=\frac{P_t}{2}$. This implies that instead of attempting to asymptotically approximate $P_R(\boldsymbol{y}_1^N)$, one can also opt to focus on doing the same for weighted Delannoy numbers. Now to proceed in this direction, we exploit the construct of generating functions, which essentially serve as a convenient way to represent an infinite sequence by displaying its coefficients as a formal power series. In regard to weighted Delannoy numbers, the corresponding generating function will be a bivariate power series, i.e., $D(q,z)=\sum_{r,s\geq 0} d_{r,s}q^r z^s$. 

We now attempt to obtain a simpler expression for $D(q,z)$ in terms of $\alpha$, $\beta$ and $\gamma$.
\begin{align}
    D(q,z)&=d_{0,0}+\sum_{r \geq 1} \beta^r q^r + \sum_{s \geq 1} \alpha^s z^s + \sum_{r,s \geq 1} d_{r,s} q^rz^s \nonumber \\
    &=1+\frac{\beta q}{1-\beta q}+ \frac{\alpha z}{1-\alpha z}+\sum_{r,s \geq 1} d_{r,s} q^rz^s. \label{eq::dxy_g}
\end{align}

The final term can be expanded by means of \eref{eq::wdel}.
\begin{align}
	&\sum_{r,s\geq 1} d_{r,s} q^rz^s
	=\sum_{r,s\geq 1}(\alpha d_{r,s-1} +\beta d_{r-1,s}+  \gamma d_{r-1, s-1}) q^rz^s \nonumber \\
    &= \alpha \! \sum_{r,s\geq 1} \! d_{r,s-1} q^rz^s + \beta  \! \sum_{r,s\geq 1} \! d_{r-1,s} q^rz^s  + \! \gamma \!\! \sum_{r,s\geq 1} \!\! d_{r-1,s-1} q^rz^s \nonumber\\
	&= \alpha z \sum_{\substack{r\geq 1\\s\geq 0}} d_{r,s} q^rz^{s} + \beta x \sum_{\substack{r\geq 0 \\ s\geq 1}} d_{r,s} q^{r}z^s  \nonumber \\
    &\qquad \quad +\gamma qz \! \sum_{r,s\geq 1} \! d_{r-1,s-1} q^{r-1}z^{s-1} \nonumber\\
	&= \alpha z \Big(\! D(q,z)-\frac{1}{1-\alpha z}\Big)\! + \! \beta q \Big( \! D(q,z)-\frac{1}{1-\beta q}\! \Big)\! \nonumber\\
    &\qquad + \! \gamma qzD(q,z) \nonumber \\
	&= (\alpha z+\beta q+\gamma qz) D(q,z) -\frac{\alpha z}{1-\alpha z}-\frac{\beta q}{1-\beta q}. \nonumber
\end{align}

Incorporating the above relation into \eref{eq::dxy_g}, we obtain
\begin{align}
    D(q,z)&=\sum_{r,s\geq 0} d_{r,s}q^r z^s =(1-\beta q - \alpha z - \gamma qz)^{-1}. \label{eq::wdel_gen}
\end{align}

The techniques in \cite{pemantle_analytic_2013, melczer_asymptotics_2020,melczer_invitation_2021} allow us to determine the asymptotic behavior of $d_{r,s}$, along a direction $\hat{\boldsymbol{r}}$ in the first quadrant. In particular, for any multivariate rational generating function, \cite[Theorem 9.5.7]{pemantle_analytic_2013} provides an expression for the asymptotics of the coefficients of its corresponding power series expansion, along any direction $\hat{\boldsymbol{r}}$. To be able to apply this theorem, we require the critical point of $D(q,z)$ along $\hat{\boldsymbol{r}}$, which essentially refers to its singular point(s) along $\hat{\boldsymbol{r}}$. In our case, these can be derived by solving $H=0$ and $rz\frac{\partial H}{\partial z}  = sq\frac{\partial H}{\partial q}$,
where ${H=1-\beta q - \alpha z - \gamma qz}$ and the point $(r,s)$ lies in the first quadrant along $\hat{\boldsymbol{r}}$. Upon doing so, we obtain the following critical points.
\begin{equation} \nonumber
	\begin{split}
	(q_1,z_1)=&\bigg(\frac{2\omega-1-\frac{1}{\eta}+\frac{\rho'}{\eta}}{2\omega \beta}, \frac{2\omega-1-\eta+\rho'}{2\omega \alpha}\bigg),  \\
	(q_2,z_2)=&\bigg(\frac{2\omega-1-\frac{1}{\eta}-\frac{\rho'}{\eta}}{2\omega \beta}, \frac{2\omega-1-\eta-\rho'}{2\omega \alpha}\bigg), 
	\end{split}
\end{equation}
where $\omega=\frac{\gamma }{\alpha \beta +\gamma }, \eta=r/s$ and $\rho'=((\eta+1)^2-4\omega \eta)^{1/2}$.

The next step involves determining the minimal point from among these critical points. This simply refers to that critical point which has the greatest impact on the asymptotics of the sequence. According to \cite{pemantle_analytic_2013}, when the concerned rational generating function is combinatorial in nature, i.e., has no negative coefficients in its power series expansion, as is the case here, it suffices to consider the critical point(s) that lie in the first quadrant alone. Furthermore, one of these critical points is guaranteed to be a minimal point. Numerical verification reveals that $(q_1,z_1)$ alone upholds this criterion, and hence, automatically qualifies as a minimal critical point. Since there are no other candidates, it is also strictly minimal. 

Now \cite[Theorem 9.5.7]{pemantle_analytic_2013} states that the asymptotic behavior of $d_{r,s}$ can be expressed as
\begin{equation}
		d_{r,s} \sim \frac{1}{\sqrt{2\pi}}q^{-r}z^{-s}\sqrt{\frac{-zH_z}{sG}}, \label{asym}
\end{equation}
where
\begin{align}
	G(q,z)&=-z^2H_z^2qH_q-zH_zq^2H_q^{{2}} \nonumber \\
	&-q^2z^2(H_z^2H_{qq}+H_q^2H_{zz}-2H_qH_zH_{qz}) \nonumber
\end{align}

The radical in \eref{asym} can be reduced to
\begin{align}
    \frac{-zH_z}{sG}&=\frac{z(\alpha+\gamma q)}{sqz(\alpha+\gamma q)(\beta+\gamma z)(\alpha z +\beta q)} \nonumber \\
    &=\frac{z}{sqz(\beta+\gamma z)(\alpha z +\beta q)} \label{term}
\end{align}

Next, we simplify the terms in the denominator of this expression.
\begin{align}
	sq_1(\beta+\gamma z_1) 
	&=\frac{s}{2\omega \beta}\big(2\omega -1 -\frac{1}{\eta}+\frac{\rho'}{\eta}\big)\big(b+\frac{\gamma }{2\omega \alpha}(2\omega -1 \nonumber\\
	&\qquad- \eta+\rho')\big) \nonumber \\
	&= \frac{s}{4\omega^2 \frac{\alpha \beta }{\alpha \beta +\gamma }\eta}\big((2\omega -1)\eta -1+\rho'\big)\big( \frac{2\omega \alpha \beta }{\alpha \beta +\gamma} \nonumber \\
	&\qquad+\frac{\gamma (2\omega -1 - \eta+\rho')}{\alpha \beta+\gamma }\big) \nonumber \\
	&= \frac{s}{4\omega^2 (1-\omega)\eta}\big((2\omega -1)\eta -1+\rho'\big)\big(2\omega \nonumber\\
	&\qquad-2\omega^2+\omega(2\omega -1 - \eta+\rho')\big) \nonumber \\
	&= \frac{s}{4\omega (1-\omega)\eta}((2\omega -1)\eta\! -1\!+\!\rho')(\rho'\!-\!\eta\!+\!1) \nonumber \\
	&= \frac{s}{2\omega}(\eta-\rho'+1). \label{t1}\\
	\alpha z_1+\beta q_1
	&=\frac{1}{2\omega}\big(4\omega-2-\eta-\frac{1}{\eta}+\rho'(\eta+\frac{1}{\eta})\big) \nonumber \\
	&= \frac{1}{2\omega\eta}\big(4\omega\eta-2\eta-\eta^2-1+\rho'(1+\eta)\big)\nonumber \\  
	&= \frac{1}{2\omega\eta}\big(-\rho'^2+\rho'(1+\eta)\big)= \frac{\rho'(\eta+1-\rho')}{2\omega\eta}. \label{t2}
\end{align}

Upon combining \eref{term}, \eref{t1} and \eref{t2}, the asymptotic expression of $d_{r,s}$ in \eref{asym} can be simplified to	${d_{r,s} \sim \sqrt\frac{\eta}{2\pi \rho' s} q_1^{-r} z_1^{-s}  \frac{2\omega}{\eta+1-\rho'}}$. Note that $\eta$ should be assigned such that it represents the direction of expected drift, which depends on the channel parameters, just like $\rho'$ and $\omega$. To proceed with the task of extracting the asymptotic behavior of $\tau_t(\boldsymbol{y})$, we may instead choose to asymptotically approximate a ratio of two weighted Delannoy numbers, say $\frac{d_{r_2,s_2}}{d_{r_1,s_1}}$ where $r_2-r_1=\delta_r<<r_1$ and $s_2-s_1=\delta_s<<s_1$. To this end, we recognize that
\begin{align}
    \lim_{r_1,s_1 \to \infty} \frac{\eta_2}{\eta_1}&= \lim_{r_1,s_1 \to \infty} \frac{r_1+\delta_r}{s_1+\delta_s} \cdot \frac{s_1}{r_1} = \lim_{r_1,s_1 \to \infty} \frac{1+\frac{\delta_r}{r_1}}{1+\frac{\delta_s}{s_1}} = 1.  \nonumber
\end{align}

Quite similarly, we also have $\lim_{r_1,s_1 \to \infty} \rho'_2 / \rho'_1= 1$. Consequently, we can write $\lim_{r_1,s_1 \to \infty} d_{r_2,s_2}/d_{r_1, s_1} = q_1^{-\delta_r} z_1^{-\delta_s}$.
This, along with \eref{eq::ptail_del}, helps us conclude that 
\begin{align}
	\lim_{r_1,s_1 \to \infty}\frac{P_{r_2}(\boldsymbol{y}_1^{s_2})}{P_{r_1}(\boldsymbol{y}_1^{s_1})}
	&= \lim_{r_1,s_1 \to \infty} \frac{d_{r_2,s_2}-\alpha d_{r_2,s_2-1}}{d_{r_1,s_1}-\alpha d_{r_1,s_1-1}} \nonumber \\
	&= \lim_{r_1,s_1 \to \infty} \frac{d_{r_2,s_2}}{d_{r_1, s_1}} \cdot \frac{1-\frac{d_{r_2,s_2-1}}{d_{r_2,s_2}}}{1-\frac{d_{r_1,s_1-1}}{d_{r_1,s_1}}} \nonumber \\
	&= \lim_{r_1,s_1 \to \infty} \frac{d_{r_2,s_2}}{d_{r_1, s_1}} \cdot \frac{1-\big(\frac{2\omega \alpha}{2\omega-1-\eta_2+\rho'_2}\big)^{-1}}{1-\big(\frac{2\omega \alpha}{2\omega-1-\eta_1+\rho'_1}\big)^{-1}} \nonumber \\
	&= \lim_{r_1,s_1 \to \infty} \frac{d_{r_2,s_2}}{d_{r_1, s_1}} = q_1^{-\delta_r} z_1^{-\delta_s}. \label{eq::asym_final}
\end{align}

By merging \eref{eq::asym_final} with \eref{eq::fano_br}, we find that
\begin{align}
	Z(\boldsymbol{v}_0^t \rightarrow \boldsymbol{v}_0^{t+1})	
	&=  \log_2 P(\boldsymbol{y}_{ct+d_{ct}+1}^{c(t+1)+d_{c(t+1)}}, d_{c(t+1)}|\boldsymbol{x}_{ct+1}^{c(t+1)}, d_{ct}) \nonumber\\
	& -b+c\log_2 q_1 +(c+d_{c(t+1)}-d_{ct}) \log_2 z_1. \nonumber 
\end{align}

\begin{rem}
    It is worth pointing out that when $P_i=P_d$, we assume the expected drift to be $0$, and thus choose to asymptotically approximate $P_r(\boldsymbol{y}_1^s)$ along the diagonal, i.e., $\eta=1$. Under these conditions, we observe
\begin{align*}
    \omega&=\frac{1-2P_i}{(1-P_i)^2}, \;\; \rho'=\frac{2P_i}{1-P_i}, \;\;
    (q_1,z_1)=(1,2).
\end{align*}
As a consequence, the asymptotic approximation of the branch decoder metric becomes
\begin{align}
	Z(\boldsymbol{v}_0^t \rightarrow \boldsymbol{v}_0^{t+1})&= \log_2 P(\boldsymbol{y}_{ct+d_{ct}+1}^{c(t+1)+d_{c(t+1)}}, d_{c(t+1)}|\boldsymbol{x}_{ct+1}^{c(t+1)}, d_{ct}) \nonumber \\
	&\quad +c+\delta-b. \nonumber
\end{align}
\end{rem}




\subsection{Proof of upper bound on complexity}

To prove that the bound in \eref{eq::cav_bound} indeed holds, we begin by attempting to simplify \eref{eq::cav_eq}. 

\subsubsection{Distribution of minimum threshold}

In this part, we derive a bound on the cumulative probability distribution of the minimum threshold, $T^*_{\mathrm{min}}$. We proceed along this direction by evaluating the cumulative probability distribution of metric values at depth $i$ along $\boldsymbol{v}_0^{*L}$, for any $\sigma<0$.
\begin{align}
	P(\mu^*_i\leq y) &= P(2^{\sigma\mu^*_i}\geq 2^{\sigma y}) \leq 2^{-\sigma y}E[2^{\sigma \mu^*_i}] \nonumber \\
	&= 2^{-\sigma y}E[2^{\sigma \sum_{j=0}^{i-1} \zeta^*_j}], \nonumber 
\end{align}
where $\zeta^*_i$ refers to the $i^{\mathrm{th}}$ branch metric along $\boldsymbol{v}_0^{*L}$. Assuming all $\zeta_i^*$'s to be independent and identically distributed yields
\begin{align}
P(\mu^*_i\leq y) &\leq2^{-\sigma y}(E[2^{\sigma \zeta^*}])^i = 2^{-\sigma y}(g_0(\sigma))^i.  \nonumber
\end{align}

Here $g_0(\sigma)$ is essentially the moment generating function of branch metrics along the correct path. Choosing a suitable $\sigma_0<0$ such that $g_0(\sigma_0)=1$, we obtain $P(\mu^*_i\leq y) \leq 2^{-\sigma_0 y}$, which also implies that
\begin{align}
	P(\mu^*_\mathrm{min}\leq y) &\leq 2^{-\sigma_0 y}. \label{eq::sig0_int}
\end{align}

Since $T_\mathrm{min}^*$ is either $0$ (when the received sequence is error-free) or any negative multiple of $\Delta$, we can further write
\begin{align}
	P(T^*_\mathrm{min}\leq y)&=\!P(T^*_\mathrm{min}\leq \Big\lfloor\frac{y}{\Delta}\Big\rfloor \Delta) = \!P(\mu^*_\mathrm{min}\leq \Big\lfloor\frac{y}{\Delta} \Big\rfloor\Delta+\Delta) \nonumber \\
	&= P(\mu^*_\mathrm{min}\leq y+\Delta) \leq 2^{-\sigma_0(y+\Delta)}. \label{eq::tmin}
\end{align}
\subsubsection{Computations in incorrect subtree}
To bound $C_\mathrm{av}$, we first recall that the number of visits to any node depends on its metric, and equals the number of unique threshold values with which it is visited. Let the initial and final threshold values with which a node $\boldsymbol{v}_0^j$ is visited, be $T_1$ and $T_2$, respectively. They are related by
\begin{equation}
	T_1=T_2+(C'(\boldsymbol{v}_0^j)-1)\Delta. \label{eq::t1}
\end{equation}
Additionally from \eref{eq::fv} we know that $T_1$ is related to $\mu(\boldsymbol{v}_0^j)$ as
\begin{equation}
	T_1 \leq \mu(\boldsymbol{v}_0^j) < T_1 + \Delta. \label{eq::t2}
\end{equation}
Since $T_1 \geq T^*_\mathrm{min}$, we can combine \eref{eq::t1} and \eref{eq::t2} into
\begin{align}
	C'(\boldsymbol{v}_0^j)&=\Big\lfloor \frac{\mu(\boldsymbol{v}_0^j)-T_1}{\Delta}\Big\rfloor+1\leq \frac{\mu(\boldsymbol{v}_0^j)-T^*_\mathrm{min}}{\Delta}+1. \nonumber
\end{align}
Using an indicator function $\phi(\boldsymbol{v}_0^j)$ defined as
\begin{equation}
\phi(\boldsymbol{v}_0^j)=
\begin{cases}
1,& \text{if node $\boldsymbol{v}_0^j$ is visited} \\
0,              & \text{else}
\end{cases} \nonumber 
\end{equation}
the previous inequality may be further refined to
\begin{align}
C'(\boldsymbol{v}_0^j)&\leq \bigg(\frac{\mu(\boldsymbol{v}_0^j)-T^*_\mathrm{min}}{\Delta}+1\bigg) \phi(\boldsymbol{v}_0^j). \nonumber 
\end{align}
Since the metric of root $\boldsymbol{v}_0^*$ is always set to zero, we may write
\begin{align}
E[C(\boldsymbol{v}^*_0) |T_\mathrm{min}^*=y] &\leq  -\frac{y}{\Delta}+1 \nonumber \\
&\hspace{-3em}+ \sum_{j=1}^{\infty} E\bigg[\sum_{\boldsymbol{v}_0^j \in \tau_{0}} C'(\boldsymbol{v}_0^j) \Big\vert T_\mathrm{min}^*=y\bigg]. \label{eq::ec1t_3}
\end{align}

We perform extensive manipulations on the latter term (see Appendix \ref{app::comp} for details) and finally arrive at the following.
\begin{align}
& E[C(\boldsymbol{v}^*_0) |T_\mathrm{min}^*=y]  \nonumber \\
&\leq -\frac{y}{\Delta}+1 +\sum_{j=1}^{\infty} E\bigg[ \sum_{\boldsymbol{v}_0^j \in \tau'_0} C'(\boldsymbol{v}_0^j) \Big\vert T^*_{\mathrm{min}}=y \bigg] \nonumber \\
&+\sum_{j=1}^{\infty} E\bigg[ \sum_{\boldsymbol{v}_0^j \in \tau^*_0} C'(\boldsymbol{v}_0^j) \Big\vert T^*_{\mathrm{min}}=y \bigg] \nonumber \\
&\leq   C_12^{-\sigma_1y}+C_22^{-(\sigma_0+\sigma_1)y}+C_3\bigg(-\frac{y}{\Delta}+1\bigg), \label{eq::al_al_m}
\end{align}

where $C_1$, $C_2$ and $C_3$ are constants for specific channel and convolutional code parameters.

We now merge \eref{eq::tmin}  and \eref{eq::al_al_m} to obtain a bound for $C_\mathrm{av}$.
\begin{align}
    &C_{\mathrm{av}}=\sum_{i=0}^{\infty} P(T^*_{\mathrm{min}}=-i \Delta) E[C(\boldsymbol{v}^*_0)|T^*_{\mathrm{min}}=-i\Delta]\nonumber\\
    &\leq C_1 \sum_{i=0}^{\infty} P(T^*_{\mathrm{min}}=-i\Delta)2^{\sigma_1 i \Delta} \nonumber\\
    &+ C_2\sum_{i=0}^{\infty} P(T^*_{\mathrm{min}}=-i\Delta)2^{(\sigma_1+\sigma_0) i \Delta}\nonumber \\
    & +C_3 \sum_{i=0}^{\infty} P(T^*_{\mathrm{min}}=-i\Delta)(i+1). \label{eq::cav_full}
\end{align}

The first summation can be further reduced as follows.
\begin{align}
    &\sum_{i=0}^{\infty} P(T^*_\mathrm{min}=-i\Delta)2^{\sigma_1 i \Delta} \nonumber\\
    &= \sum_{i=0}^{\infty} (P(T^*_\mathrm{min}\leq -i\Delta) -P(T^*_\mathrm{min}\leq -(i+1)\Delta))2^{\sigma_1 i \Delta} \nonumber \\
    &=\sum_{i=0}^{\infty} P(T^*_\mathrm{min}\leq -i\Delta)2^{\sigma_1 i \Delta} -\!\! \sum_{i=0}^{\infty} \! P(T^*_\mathrm{min}\leq -i\Delta)2^{\sigma_1 (i-1) \Delta} \nonumber\\
    &\qquad+2^{-\sigma_1 \Delta} P(T^*_{\mathrm{min}} \leq 0) \nonumber\\
    &=(1-2^{-\sigma_1 \Delta})\sum_{i=0}^{\infty} P(T^*_{\mathrm{min}}\leq -i \Delta)2^{\sigma_1 i \Delta}+2^{-\sigma_1 \Delta} \nonumber \\
    &\leq (1-2^{-\sigma_1 \Delta}) \sum_{i=0}^{\infty} 2^{(i-1)\sigma_0\Delta+\sigma_1 i \Delta}+2^{-\sigma_1 \Delta} \nonumber \\
    &= (1-2^{-\sigma_1 \Delta})2^{-\sigma_0 \Delta} \sum_{i=0}^{\infty} 2^{(\sigma_0+\sigma_1)i \Delta}+2^{-\sigma_1 \Delta}, \label{eq::sum1}
\end{align}
where the penultimate relation is due to \eref{eq::tmin}. Similarly, the second summation in \eref{eq::cav_full} can be simplified to
\begin{align}
    &\sum_{i=0}^{\infty} P(T^*_{\mathrm{min}}=-i \Delta)2^{(\sigma_0+\sigma_1)i\Delta} \nonumber \\
    &=(1-2^{-(\sigma_0+\sigma_1)i \Delta}) \sum_{i=0}^{\infty} P(T^*_{\mathrm{min}} \leq - i \Delta)2^{(\sigma_0+\sigma_1)i\Delta} \nonumber \\
    &\qquad+2^{-(\sigma_0+\sigma_1)\Delta} \nonumber\\
    &= (1-2^{-(\sigma_0+\sigma_1)i \Delta}) \sum_{i=0}^{\infty} 2^{(i-1)\sigma_0\Delta+(\sigma_0+\sigma_1)i\Delta} +2^{-(\sigma_0+\sigma_1)\Delta}.  \label{eq::sum2}
\end{align}

The final summation in \eref{eq::cav_full} is dealt with by undertaking a similar approach.
\begin{align}
    &\sum_{i=0}^{\infty} P(T^*_{\mathrm{min}}=-i \Delta)(i+1) \nonumber \\
    &=\sum_{i=0}^{\infty} P(T^*_{\mathrm{min}} \leq -i \Delta)(i+1) -\sum_{i=1}^{\infty} P(T^*_{\mathrm{min}} \leq -i \Delta)i \nonumber \\
    &= \sum_{i=0}^{\infty} P(T^*_{\mathrm{min}} \leq -i \Delta) \leq \frac{2^{-\sigma_0 \Delta}}{1-2^{\sigma_0 \Delta}}. \label{eq::sum3}
\end{align}

The final inequality holds since $\sigma_0<0$. We note that \eref{eq::sum1} and \eref{eq::sum2} also converge if $\sigma_0+\sigma_1<0$. By assuming that this is indeed true and incorporating \eref{eq::sum1}, \eref{eq::sum2} and \eref{eq::sum3} into \eref{eq::cav_full}, we arrive at \eref{eq::cav_bound}.

\subsection{Computations in incorrect subtree} \label{app::comp}
To decompose the term $\sum_{j=1}^{\infty} E[\sum_{\boldsymbol{v}_0^j \in \tau_{0}} C'(\boldsymbol{v}_0^j)| T_\mathrm{min}^*=y]$ as required in (\ref{eq::ec1t_3}), we subdivide $\tau_0$ into two subtrees: $\tau_{0}'$, that contains nodes which hypothesize a wholly inaccurate message sequence, and $\tau_{0}^*$, which consists of nodes that follow the true message sequence, at least initially, but suggest a different drift sequence. We depict this in Fig \ref{fig::ct}.

\subsubsection{On incorrect paths}

We proceed by characterizing the distribution of node metrics, or equivalently branch metrics, in $\tau'_{0}$. Given their dependence on relative drift changes, these branch metrics are treated in a drift-specific manner. Let $Z'(\delta)$ describe the distribution of branch metrics in $\tau'_{0}$ for a specific drift change $\delta$. Also, let $i_\mathrm{max}$ and $d_\mathrm{max}$ denote the maximum allowable insertions and deletions over a single block. We can bound the number of visits to nodes in $\tau'_{0}$ at depth $1$, as 
\begin{align}
&E\bigg[\sum_{\boldsymbol{v}_0^1 \in \tau'_{0}} \bigg(\frac{\mu(\boldsymbol{v}_0^1)-y}{\Delta}+1\bigg) \phi(\boldsymbol{v}_0^1) \bigg| T_\mathrm{min}^*=y\bigg] \nonumber \\
&=E\bigg[\sum_{\boldsymbol{v}_0^1 \in \tau'_{0}} \bigg(\frac{Z(\boldsymbol{v}_0\rightarrow\boldsymbol{v}_0^1)-y}{\Delta}+1\bigg) \phi(\boldsymbol{v}_0^1) \bigg| T_\mathrm{min}^*=y\bigg] \nonumber\\
&=(2^b-1)\sum_{\delta=-d_\mathrm{max}}^{i_\mathrm{max}}\sum_{z_1\geq y} \bigg(\frac{z_1-y}{\Delta}+1\bigg)P(Z'(\delta)=z_1) \nonumber \\
&=(2^b-1)\lambda \sum_{z_1\geq y} \bigg(\frac{z_1-y}{\Delta}+1\bigg)P(\zeta'=z_1), \label{eq::t0'1}
\end{align}
where $\lambda=i_\mathrm{max}+d_\mathrm{max}+1$ and the variable $\zeta'$ uniformly combines the probability distributions of all the drift-specific random variables, i.e.,
\begin{equation}
    P(\zeta'=z)=\frac{1}{\lambda} \sum_{\delta=-d_{\mathrm{max}}}^{i_{\mathrm{max}}} P(Z'(\delta)=z). \label{eq::pzpp}
\end{equation}

By assuming that branch metrics across different levels are independently distributed, we arrive at the following generalization of \eref{eq::t0'1} for higher depths
\begin{align}
&E\bigg[\sum_{\boldsymbol{v}_0^j \in \tau'_{0}}  \bigg(\frac{\mu(\boldsymbol{v}_0^j)-y}{\Delta}+1\bigg) \phi(\boldsymbol{v}_0^j) \bigg| T_\mathrm{min}^*=y\bigg] \nonumber \\
&=(2^b-1)2^{b(j-1)} \lambda^j\sum_{\mu \geq y} \bigg(\frac{\mu -y}{\Delta}+1\bigg) f_j(y,\mu), \label{eq::end1}
\end{align}
where $f_j(y,\mu)= P(\mu'_1 \geq y,\cdots \mu'_{j-1} \geq y, \mu'_j=\mu)$. This quantity refers to the probability that during a random walk along the code tree, the initial $j-1$ node metrics $\mu'_1, \ldots \mu'_{j-1}$ lie above threshold $y$, and the final metric $\mu'_{j}=\mu$, i.e.,
\begin{equation}
f_j(y,\mu)= P(\mu'_1 \geq y,\cdots \mu'_{j-1} \geq y, \mu'_j=\mu). \label{eq::h1}
\end{equation}

The net decoding effort expended in $\tau'_0$ is thus
\begin{align}
    &\sum_{j=1}^{\infty} E\big[\!\!\sum_{\boldsymbol{v}_0^j \in \tau'_0} \!\!\! C'(\boldsymbol{v}_0^j)|T^*_{\mathrm{min}}=y \big] =\! (1-2^{-b}) \! \sum_{j=1}^{\infty} 2^{b' j} \!\sum_{\mu \geq y} f_j(y,\mu) \nonumber \\
    &\qquad+(1-2^{-b})\Delta^{-1} \sum_{j=1}^{\infty} 2^{b' j} \sum_{\mu \geq y} (\mu -y)f_j(y,\mu), \label{eq::tau_p_0}
\end{align}
where $b'=b+\log_2 \lambda$. To unravel the former summation in the previous equation, we make use of the following equality.
\begin{equation}
\sum_{t=1}^{\infty} \sum_{z < y} f_t(y,z) =1. \label{eq::walds}
\end{equation}

This relation implies that an infinite random walk in $\tau_{0}$, and in turn also $\tau'_0$, will eventually fall below a finite $T^*_\mathrm{min}$, provided that $E[\zeta']<0$, which is indeed true for reasonable channel parameters as confirmed by numerical verification. This is crucial in ensuring that Fano's decoder is unlikely to pick the wrong successor at any step. To apply this for the simplification of the initial term in (\ref{eq::tau_p_0}), we recognize that $\sum_{z \geq y} f_j(y,z)$ can be interpreted as the probability that during an infinite random walk, the metric of the node at depth $j$, i.e. $z$, has not fallen below the barrier $y$. From another perspective, this quantity can also be seen as the probability that during a random walk, the node metric at a depth beyond $j$, falls below the barrier $y$. Mathematically, we express this equivalence as	$\sum_{z \geq y} f_j(y,z) =\sum_{t=j+1}^{\infty} \sum_{\mu < y} f_t(y, \mu)$. Upon applying this to the first summation in (\ref{eq::tau_p_0}), we obtain
\begin{align}
	&\sum_{j=1}^{\infty} 2^{b'j} \sum_{\mu \geq y} f_j(y, \mu) \nonumber =\sum_{j=1}^{\infty} 2^{b'j} \sum_{t=j+1}^{\infty} \sum_{\mu<y} f_t(y, \mu) \nonumber \\
	&= \sum_{j=1}^{\infty} \frac{2^{b'j}-2{b'(j-1)}}{1-2^{-b'}} \sum_{t=j+1}^{\infty} \sum_{\mu<y} f_t(y, \mu) \nonumber \\
	&=(1-2^{-b'})^{-1} \bigg(\sum_{j=1}^{\infty} 2^{b'j} \sum_{t=j+1}^{\infty} \sum_{\mu<y} f_t(y,\mu) \nonumber\\
	&\qquad-\sum_{j=1}^{\infty} 2^{b'(j-1)} \sum_{t=j+1}^{\infty} \sum_{\mu<y} f_t(y,\mu)\bigg) \nonumber \\
	&=(1-2^{-b'})^{-1} \bigg(\sum_{j=1}^{\infty} 2^{b'j} \sum_{t=j+1}^{\infty} \sum_{\mu<y} f_t(y,\mu) \nonumber\\
	&\qquad-\sum_{j=0}^{\infty} 2^{b'j} \sum_{t=j+2}^{\infty} \sum_{\mu<y} f_t(y,\mu)+\sum_{t=1}^{\infty} \sum_{\mu<y}f_t(y, \mu)-1\bigg) \nonumber \\
	&=(1-2^{-b'})^{-1} \bigg(\sum_{j=0}^{\infty} 2^{b'j} \sum_{t=j+1}^{\infty} \sum_{\mu<y} f_t(y,\mu) \nonumber\\
	&\qquad\qquad -\sum_{j=0}^{\infty} 2^{b'j} \sum_{t=j+2}^{\infty} \sum_{\mu<y} f_t(y,\mu)-1\bigg) \nonumber \\
	&=(1-2^{-b'})^{-1} \big(\sum_{j=0}^{\infty}2^{b'j} \sum_{\mu<y} f_{j+1}(y, \mu)-1\big). \label{eq::si_p1}
\end{align}

To further reduce this quantity, we utilize a tilted probability assignment for branch metrics along incorrect paths.
\begin{equation}
	P(\zeta'_{\sigma}=z)=\frac{2^{\sigma z }P(\zeta'=z)}{g_1(\sigma)}, \label{eq::p_sig}
\end{equation}
where $g_1(\sigma)$ denotes the moment generating function of branch metrics along an incorrect path, i.e.,
\begin{align}
    g_1(\sigma)&= E[2^{\sigma \zeta'}] =\sum_{z} P(\zeta'=z)2^{\sigma z}. \label{eq::g_1_def}
\end{align}

The probability distribution in \eref{eq::p_sig} is clearly valid since $P(\zeta'_{\sigma}=z)\geq 0$ for all $z$, and
\begin{align}
	\sum_z P(\zeta'_{\sigma}=z)&={g_1(\sigma)^{-1}} \sum_z 2^{\sigma z} P(\zeta'=z) =1. \nonumber
\end{align}

We now consider a random walk $p'_{\sigma}$ through the incorrect subtree $\tau'_0$, with the associated branch metrics being denoted by the random variables $\zeta'_{0,\sigma}, \zeta'_{1,\sigma}, \ldots$. These variables will be guided by the probability distribution in (\ref{eq::p_sig}), and we may write $P(\zeta'_{i,\sigma}=a)=P(\zeta'_{i}=a)2^{\sigma a}{g_1(\sigma)^{-1}} $,
where the random variables $\zeta'_0, \zeta'_1, \ldots$ denote the branch metrics encountered during a random walk in $\tau'_0$, guided by the probability distribution in (\ref{eq::pzpp}). Thus, analogously to (\ref{eq::h1}), we can define $f_{\sigma, j}(y,\mu)=P(\mu'_{\sigma,1} \geq y, \ldots, \mu'_{\sigma,j-1} \geq y, \mu'_{\sigma,j}=\mu)$. 

We now consider a set $\mathcal{Z}^{(j)}(y) \in \mathbb{R}^j$ containing all vectors $\boldsymbol{z}=(z_1, \ldots, z_j)$, that for a specific $T^*_{\mathrm{min}}=y$, satisfy $\sum_{i=1}^{k} z_i \geq y$ for all $1 \leq k \leq j$.
This relation can be used to restate $f_{\sigma, j}(y,\mu)$ as
\begin{align}
    &f_{\sigma, j}(y, \mu) \nonumber\\
    &= \hspace{-1.2em}\sum_{\substack{\boldsymbol{z}_1^{(j-1)} \\ \in \boldsymbol{\mathcal{Z}}^{(j-1)}(y)}} \hspace{-1.2em}  P(\zeta'_{1,\sigma}=z_1, \ldots, \zeta'_{j-1,\sigma}=z_{j-1}, \zeta'_{j,\sigma}=\mu-\sum_{i=1}^{j-1}z_i) \nonumber \\
    &= \sum_{\substack{\boldsymbol{z}_1^{(j-1)} \\ \in \boldsymbol{\mathcal{Z}}^{(j-1)}(y)}}  \big(\prod_{i=1}^{j-1} P(\zeta'_{i,\sigma}=z_i)\big) P(\zeta'_{j,\sigma}=\mu-\sum_{i=1}^{j-1}z_i) \nonumber \\
    &= \hspace{-1.2em}  \sum_{\substack{\boldsymbol{z}_1^{(j-1)} \\ \in \boldsymbol{\mathcal{Z}}^{(j-1)}(y)}} \hspace{-1.5em} \frac{\big(\prod_{i=1}^{j-1} P(\zeta'_{i}=z_i) 2^{\sigma z_i} \big)}{{g_1(\sigma)^j}}  P\Big(\zeta'_{j}=\! \mu- \! \sum_{i=1}^{j-1}\! z_i\Big)\frac{2^{\sigma \mu}}{2^{\sigma \sum_{i=1}^{j-1}z_i}} \nonumber \\
    &= \frac{2^{\sigma \mu}}{g_1(\sigma)^{j}} \hspace{-1.2em} \sum_{\substack{\boldsymbol{z}_1^{(j-1)} \\ \in \boldsymbol{\mathcal{Z}}^{(j-1)}(y)}} \hspace{-1.2em} P(\zeta'_1=z_1, \ldots, \zeta'_{j-1}=z_{j-1}, \zeta'_j=\mu - \sum_{i=1}^{j-1}z_i) \nonumber \\
    &=g_1(\sigma)^{-j} 2^{\sigma \mu} f_{0,j}(y, \mu). \nonumber
\end{align}

If we choose $\sigma$ such that $E[\zeta'_{\sigma}]=\sum_{z'} P(\zeta'_{\sigma}=z')$ remains negative as it is for an unbiased random walk dictated by \eref{eq::pzpp}, we can rewrite \eref{eq::walds} as follows.
\begin{equation}
    \sum_{t=1}^{\infty} \sum_{\mu < y} f_{\sigma, t}(y,\mu)=\sum_{t=1}^{\infty}\sum_{\mu < y} f_{0,t}(y,\mu)2^{\sigma \mu} g_1(\sigma)^{-t}=1. \nonumber
\end{equation}

Picking a $\sigma_1>0$ such that $g_1(\sigma_1)=2^{-b'}$, the previous equation transforms into
\begin{equation}
    \sum_{t=1}^{\infty} \sum_{\mu < y} f_{\sigma_1, t} (y,\mu) = \sum_{t=1}^{\infty}\sum_{\mu < y} f_{0,t}(y,\mu)2^{\sigma_1 \mu} 2^{b' t}=1. \label{eq::f_sig_eq}
\end{equation}

Now recall the definition of $f_j(y,\mu)$ in \eref{eq::h1} and note that
\begin{equation}
    \sum_{\mu<y} \hspace{-0.2em} f_{0,t}(y,\mu)2^{\sigma_1 \mu}\hspace{-0.3em}=\hspace{-0.4em}\sum_{\mu<y} \!P(\mu'_1\geq y, \ldots, \mu'_{t-1} \geq y, \mu'_t=\mu)2^{\sigma_1 \mu}. \nonumber
\end{equation}

Now given that $\mu'_{t-1} \geq y$, it must hold that $\mu'_t=\mu'_{t-1}+\zeta'_t \geq y+z'_{\mathrm{min}}$, where $z'_{\mathrm{min}}<0$ denotes the minimum value that $\zeta'$ might assume. As a consequence, the preceding expression can be lower-bounded as follows.
\begin{equation}
    \sum_{\mu<y} f_{0,t}(y,\mu)2^{\sigma_1 \mu} \geq 2^{\sigma_1(y+z'_{\mathrm{min})}} \sum_{\mu < y} f_{0,t}(y,\mu). \label{eq::pf_lb}
\end{equation}

By incorporating \eref{eq::pf_lb} into \eref{eq::f_sig_eq}, we obtain
\begin{align}
    2^{\sigma_1(y+z'_{\mathrm{min}})} \sum_{t=1}^{\infty}2^{b' t} \sum_{\mu<y} f_{0,t}(y,\mu) \! &\leq \! \sum_{t=1}^{\infty} \sum_{\mu<y} f_{0,t}(y,\mu)2^{\sigma_1 \mu+b' t} \nonumber \\
    &=1 \nonumber\\
    \implies \sum_{t=1}^{\infty} 2^{b' t} \sum_{\mu<y} f_{0,t}(y,\mu) &\leq 2^{-\sigma_1(y+z'_{\mathrm{min}})}. \label{eq::zmin_stuff}
\end{align}

Applying this to \eref{eq::si_p1}, we can arrive at
\begin{align}
    \sum_{j=1}^{\infty} \sum_{\mu \geq y} f_j(y,\mu) &= (1-2^{-b'})^{-1} \big( \sum_{j=0}^{\infty} 2^{b' j} \sum_{\mu<y} f_{j+1} (y,\mu)\! -1\big) \nonumber \\
    &= \! (1-2^{-b'})^{-1} \big( \sum_{j=1}^{\infty} 2^{b' (j-1)}\! \sum_{\mu<y}\! f_{j} (y,\mu)\! -\!1\big) \nonumber \\
    & \leq \frac{2^{-\sigma_1(y+z'_{\mathrm{min}})-b'}-1}{1-2^{-b'}}. \label{eq::UB_fj_inf}
\end{align}

Hence, we can establish an upper bound on the number of nodes in $\tau'_0$ that do not fall below the threshold $y$.
\begin{equation}
    (1-2^{-b})\sum_{j=1}^{\infty} 2^{b' j} \sum_{\mu \geq y}\! f_j(y,\mu) \! \leq \frac{1-2^{-b}}{1-2^{-b'}}(2^{-\sigma_1(y+z'_{\mathrm{min}})-b'}\!-\!1). \nonumber
\end{equation}

Now to resume the task of bounding the decoding effort spent on paths in $\tau'_0$, we now focus on the latter summation in \eref{eq::tau_p_0}. To proceed along this direction, we observe that differentiating \eref{eq::f_sig_eq} leads to
\begin{align}
    \sum_{j=1}^{\infty} \! \sum_{\mu<y} \! f_{0,j}(y,\mu)2^{\sigma \mu} \big(\mu  g_1(\sigma)^{-j} \mathrm{ln} \; 2 - j g_1(\sigma)^{-j-1} g'_1(\sigma)\big)\!=\!0. \label{eq::fj_diff}
\end{align}

From \eref{eq::g_1_def}, we also note that
\begin{equation}
    E[\zeta']=\left. \frac{g_1'(\sigma)}{\ln 2} \right\vert_{\sigma=0}. \label{eq::ezeta_p}
\end{equation}

This can be used to rewrite \eref{eq::fj_diff} as
\begin{align}
    &\sum_{j=1}^{\infty} \sum_{\mu<y} f_{0,j}(y,\mu)\mu 2^{\sigma \mu} g_1(\sigma)^{-j} \nonumber \\
    &=\sum_{j=1}^{\infty} \sum_{\mu<y} f_{0,j}(y,\mu) 2^{\sigma \mu} j g_1(\sigma)^{-j-1} E[\zeta]. \nonumber
\end{align}

By evaluating the preceding equality at $\sigma=0$, we get
\begin{align}
    \sum_{j=1}^{\infty} \sum{\mu<y} f_{0,j}(y,\mu) \mu &= E[\zeta'] \sum_{j=1}^{\infty}j\sum_{\mu<y} f_{0,j}(y,\mu) \nonumber \\
    E[\mu'_N|T^*_{\mathrm{min}}=y] &= E[\zeta'] E[N|T^*_{\mathrm{min}}=y], \nonumber
\end{align}
where $N$ is a random variable indicating the depth in the joint channel and code tree, where the node metric is expected to drop below threshold $T^*_{\mathrm{min}}=y$ for the first time. Now since we have $y+z'_{\mathrm{min}} \leq E[\mu'_N|T^*_{\mathrm{min}}=y] \leq y$, 
\begin{align}
    y &\geq E[\zeta'] E[N|T^*_{\mathrm{min}}=y] =E[\zeta'] \sum_{j=1}^{\infty} j \sum_{\mu<y} f_{0,j}(y,\mu). \label{eq::y_lb}
\end{align}

This is equivalent to
\begin{align}
    y-\mu \geq E[\zeta'] \sum_{j=1}^{\infty} j \sum_{x<y-\mu} f_{j} (y-\mu,x) \nonumber \\
    \implies \mu-y \leq -E[\zeta']\sum_{j=1}^{\infty} j \sum_{x<y-\mu} f_{j} (y-\mu,x). \nonumber
\end{align}

Hence, the latter term in \eref{eq::tau_p_0} can be bounded.
\begin{align}
    &\sum_{j=1}^{\infty} 2^{b' j} \sum_{\mu \geq y} (\mu-y) f_j(y,\mu)  \nonumber \\
    & \leq -E[\zeta'] \sum_{j=1}^{\infty} 2^{b' j} \sum_{\mu \geq y} \sum_{t=1}^{\infty} t \sum_{x<y-\mu} f_t (y-\mu,x) f_j(y,\mu) \nonumber \\
    &= -E[\zeta'] \sum_{j=1}^{\infty} 2^{b' j} \sum_{\mu \geq y} \sum_{t=1}^{\infty} t f_j (y,\mu) \sum_{x<y-\mu} f_t(y-\mu,x) \nonumber \\
    &= -E[\zeta'] \sum_{j=1}^{\infty} 2^{b' j} \sum_{t=1}^{\infty} t \sum_{\mu \geq y} f_j(y,\mu) \sum_{x<y-\mu} f_t(y-\mu,x). \label{eq::tau_p_sum2}
\end{align}

To further simplify $\sum_{\mu \geq y} f_j(y,\mu) \sum_{x<y-\mu} f_t(y-\mu,x)$, we try to gain an insight to it by considering the following.
\begin{itemize}
    \item $f_j(y,\mu)$ signifies that at depth $j$, the node metric is $\mu$ and the barrier $y$ has not been crossed, i.e., $\mu \geq y$.
    \item The term $\sum_{x<y-\mu} f_t(y-\mu,x)$ signifies the probability that a random walk starting from the root with metric $0$, crosses the barrier $y-\mu$ at depth $t$; or equivalently, the probability that a random walk where the starting node has metric $\mu$, crosses the barrier $y$ at {depth $t$}.
\end{itemize}

By combining these observations, we can interpret the term $f_j(y,\mu)f_t(y-\mu, x)$ as the probability that during a random walk, metrics at depths $j$ and $j+t$ are $\mu$ and $x+\mu$ respectively. Furthermore, the term $\sum_{\mu \geq y} f_j(y,\mu)\sum_{x<y-\mu} f_t(y-\mu, x)$ can be viewed as the probability that during a random walk, metric at depth $j$ stays above the barrier $y$, but the metric at depth $j+t$ falls below $y$. More concisely, we may write $\sum_{\mu \geq y} f_j(y,\mu) \sum_{x<y-\mu} f_t(y-\mu,x)=\sum_{\mu<y} f_{j+t}(y,\mu)$,
since the definition in \eref{eq::h1} already guaranteed that the nodes at depths less than $j+t$ have metrics lying above the barrier $y$. We apply this simplification to \eref{eq::tau_p_sum2} and obtain
\begin{align}
    &\sum_{j=1}^{\infty}2^{b' j} \sum_{\mu \geq y} (\mu-y) f_j(y,\mu) \nonumber \\
    &=-E[\zeta'] \sum_{j=1}^{\infty} 2^{b' j} \sum_{t=1}^{\infty}t \sum_{\mu<y} f_{j+t}(y,\mu) \nonumber \\
    &=-E[\zeta'] \sum_{j=1}^{\infty} 2^{b' j} \sum_{k-j=1}^{\infty} (k-j) \sum_{\mu<y} f_k(y,\mu) \nonumber \\
    &=-E[\zeta'] \sum_{k=1}^{\infty} \sum_{j=1}^{k} (k-j)2^{b' j} \sum_{\mu<y} f_k(y,\mu) \nonumber \\
    &= -E[\zeta'] \sum_{k=1}^{\infty} \sum_{\mu<y} f_k(y,\mu) \sum_{j=1}^{k} (k-j)2^{b' j}. \label{eq::tau0p_sum2}
\end{align}
We can further expand $\sum_{j=1}^{k} (k-j)2^{b' j}$ as follows.
\begin{align}
    \sum_{j=1}^{k} (k-j)2^{b' j}&=k\sum_{j=1}^{k} 2^{b' j} - \sum_{j=1}^{k} j 2^{b' j} \nonumber \\
    &=k2^{b'} \frac{2^{b' k}-1}{2^{b'}-1}+\frac{2^{b'(k+1)}}{(2^{b'}-1)^2}-\frac{2^{b'}}{(2^{b'}-1)^2} \nonumber \\
    &=\frac{-k2^{b'}}{2^{b'}-1}+\frac{2^{b'(k+1)}}{(2^{b'}-1)^2}-\frac{2^{b'}}{(2^{b'}-1)^2}. \nonumber
\end{align}
Incorporating this result in \eref{eq::tau0p_sum2},
\begin{align}
    &\sum_{j=1}^{\infty}2^{b'j} \sum_{\mu \geq y} (\mu-y) f_j(y,z)  \leq \frac{2^{b'}}{2^{b'}-1}E[\zeta'] \sum_{k=1}^{\infty} k \sum_{\mu<y}f_k(y,\mu) \nonumber \\
    &\hspace{28mm}-\frac{2^{b'}}{(2^{b'}-1)^2} E[\zeta'] \sum_{k=1}^{\infty} 2^{b' k}\sum_{\mu<y} f_k (y,\mu) \nonumber \\
    &\hspace{28mm} +\frac{2^{b'}}{(2^{b'}-1)^2} E[\zeta'] \sum_{k=1}^{\infty} \sum_{\mu<y} f_k(y,\mu) \nonumber \\
    &=\frac{2^{b'}}{2^{b'}-1}E[\zeta'] \sum_{k=1}^{\infty} k \sum_{\mu<y}f_k(y,\mu)+\frac{2^{b'}}{(2^{b'}-1)^2} E[\zeta'] \nonumber \\
    &\qquad-\frac{2^{b'}}{(2^{b'}-1)^2} E[\zeta'] \sum_{k=1}^{\infty} 2^{b' k} \sum_{\mu<y}f_k(y,\mu). \nonumber
\end{align}

By exploiting \eref{eq::zmin_stuff}, \eref{eq::ezeta_p} and \eref{eq::y_lb}, we can reduce the preceding equation to
\begin{align}
    \sum_{j=1}^{\infty}2^{b' j} \sum_{\mu \geq y} (\mu-y) f_j(y,z) & \leq \frac{2^{b'}y}{2^{b'}-1}-\frac{2^{b'-\!\sigma_1(y+z'_{\mathrm{min}})}}{(2^{b'}-1)^2} E[\zeta'] \nonumber \\
    &\hspace{-7mm}= \frac{2^{b'}}{2^{b'}-1} \bigg(y-\frac{2^{-\sigma_1(y+z'_{\mathrm{min}})}}{2^{b'}-1} E[\zeta'] \bigg). \label{eq::UB_sum2_tau0p}
\end{align}

We now combine \eref{eq::UB_fj_inf} and \eref{eq::UB_sum2_tau0p} in order to bound \eref{eq::tau_p_0}.
\begin{align}
    &\sum_{j=1}^{\infty}2^{b' j} \sum_{\mu \geq y} \bigg( \frac{\mu-y}{\Delta} +1\bigg) f_j(y,\mu) \nonumber \\
    &\leq \! \frac{2^{-\sigma_1(y+z'_{\mathrm{min}})-b'}-1}{1-2^{-b'}}\!+\!\frac{\Delta^{-1}}{1-2^{-b'}} \bigg( y-\frac{2^{-\sigma_1(y+z'_{\mathrm{min}})}}{2^{b'}-1}E[\zeta'] \bigg) \nonumber \\
    &=C'_1 2^{-\sigma_1 y} - \frac{1-\Delta^{-1}y}{1-2^{-b'}}, \label{eq::bound_tau0p}
\end{align}
where $C'_1=\frac{2^{-b'}}{1-2^{-b'}}-\frac{\Delta^{-1}2^{-b'}E[\zeta']}{(1-2^{-b'})^2}$. Since $E[\zeta']$ is typically negative, $C'_1$ is evidently a positive quantity. By applying \eref{eq::bound_tau0p} to \eref{eq::tau_p_0}, we are finally able to obtain the following bound on the total decoding complexity associated with nodes in $\tau'_0$.
\begin{align}
    &\sum_{j=1}^{\infty} E[\sum_{\boldsymbol{v}_0^j \in \tau'_0} \bigg(\frac{\mu(\boldsymbol{v}_0^j)-y}{\Delta}+1 \bigg)\phi(\boldsymbol{v}_0^j) \bigg| T^*_{\mathrm{min}}=y] \nonumber \\
    & \leq (1-2^{-b})C'_12^{-\sigma_1 y}-\frac{1-2^{-b}}{1-2^{-b'}} \bigg(-\frac{y}{\Delta} +1\bigg). \label{eq::UB_tau0p}
\end{align}
\subsubsection{On partially true paths}
In a similar vein, we also analyze the distribution of node metrics in $\tau_0^*$. Evidently, branch metrics in $\tau^*_{0}$ follow a different probability distribution than $\tau'_{0}$, since the predicted block output is not necessarily independent of the received frame, unlike the previous case. Hence, we let a random variable $Z^*(\delta)$ characterize the distribution of metrics along such branches, for a specific drift change $\delta$. We also note that multiple paths hypothesizing the same message sequence as $\boldsymbol{v}_0^{*L}$ but with alternate drift sequences, may yield the same node metrics as in $\boldsymbol{v}_0^{*L}$. For instance, an all-zero codeword with a nonzero final drift could correspond to multiple drift sequences that seem equally likely. However, for well-behaved codewords\footnote{Only few equally likely drift sequences exist, after one insertion or deletion.}, such alternate paths are far more infrequent. Additionally, we observe that for higher depths, most of these alternate drift paths lead to a sizeable shift between the received and hypothesized sequences, making them appear random with respect to each other. Thus, we can reasonably assume that all branch metrics in $\tau^*_0$ beyond depth $1$ can be described by $\zeta'$. Armed with this assumption and (\ref{eq::end1}), we bound the average number of visits to nodes in $\tau^*_0$.
\begin{align}
&\sum_{\boldsymbol{v}_0^j \in \tau^*_{0}} E\bigg[ \bigg(\frac{\mu(\boldsymbol{v}_0^j)-y}{\Delta}+1\bigg) \phi(\boldsymbol{v}_0^j) \bigg| T_\mathrm{min}^*=y\bigg] \nonumber \\
&= 2^{b'(j-1)}\hspace{-1em}\sum_{\substack{\delta_1^*, \delta_1=-d_\mathrm{max}\\ \delta_1 \neq \delta^*_1}}^{i_\mathrm{max}}\hspace{-1.5em} P(\delta_1^*)\sum_{z_1\geq y}P(Z_1^*(\delta_1)=z_1)\hspace{-0.7em}\sum_{\mu \geq y-z_1} \bigg(1 \nonumber \\
&+ \frac{\mu - (y-z_1)}{\Delta}\bigg) f_{j-1}(y-z_1,\mu). \nonumber
\end{align}
The previous relation can be rewritten as
\begin{align}
    &\sum_{\boldsymbol{v}_0^{j+1} \in \tau^*_0} E\bigg[\bigg(\frac{\mu(\boldsymbol{v}_0^{j+1})-y}{\Delta}+1\bigg)\phi(\boldsymbol{v}_0^j)\bigg| T^*_\mathrm{min}=y\bigg] \nonumber\\
    &=2^{b' j} \sum_{\substack{\delta^*_1, \delta_1=-d_{\mathrm{max}}\\\delta_1 \neq \delta^*_1}} P(\delta^*_1) \sum_{z_1\geq y} P(Z^*_1(\delta_1)=z_1) \sum_{\mu \geq y-z_1} \bigg(1 \nonumber \\
    & \qquad \qquad \qquad+\frac{\mu-(y-z_1)}{\Delta}\bigg)f_j(y-z_1,\mu). \nonumber
\end{align}

We can use this equality to restate the average decoding effort expended on all nodes in $\tau^*_0$ in the following manner.
\begin{align}
    &\sum_{j=1}^{\infty} E \bigg[ \sum_{\boldsymbol{v}_0^j \in \tau^*_0} \bigg( \frac{\mu(\boldsymbol{v}_0^j)-y}{\Delta}+1 \bigg) \phi(\boldsymbol{v}_0^j) \bigg|T^*_{\mathrm{min}}=y \bigg] \nonumber \\
    &=E\bigg[ \sum_{\boldsymbol{v}_0^1 \in \tau^*_0} \bigg( \frac{\mu(\boldsymbol{v}_0^1)-y}{\Delta}+1\bigg) \phi(\boldsymbol{v}_0^1) \bigg| T^*_{\mathrm{min}}=y \bigg] \nonumber \\
    &+\sum_{j=2}^{\infty} E\bigg[ \sum_{\boldsymbol{v}_0^j \in \tau^*_0} \bigg( \frac{\mu(\boldsymbol{v}_0^j)-y}{\Delta}+1 \bigg) \phi(\boldsymbol{v}_0^j) \bigg| T^*_{\mathrm{min}}=y\bigg] \nonumber \\
    &=\sum_{\delta^*_1} P(\delta^*_1) \sum_{\substack{\delta_1=-d_{\mathrm{max}}\\ \delta_1 \neq \delta^*_1}}^{i_{\mathrm{max}}} \sum_{z_1\geq y}\bigg( \frac{z_1-y}{\Delta}+1\bigg)P(Z^*(\delta_1)=z_1) \nonumber \\
    &+\sum_{j=1}^{\infty} E\bigg[ \sum_{\boldsymbol{v}_0^{j+1} \in \tau^*_0} \bigg( \frac{\mu(\boldsymbol{v}_0^{j+1})-y}{\Delta}+1 \bigg) \phi(\boldsymbol{v}_0^{j+1}) \bigg| T^*_{\mathrm{min}}=y\bigg]. \nonumber
\end{align}

Reusing \eref{eq::UB_tau0p} to simplify the previous equation, we get
\begin{align}
    &\sum_{j=1}^{\infty} E \bigg[ \sum_{\boldsymbol{v}_0^j \in \tau^*_0} \bigg( \frac{\mu(\boldsymbol{v}_0^j)-y}{\Delta}+1 \bigg) \phi(\boldsymbol{v}_0^j) \bigg|T^*_{\mathrm{min}}=y \bigg] \nonumber \\
    & \leq \sum_{\substack{\delta^*_1, \delta_1=-d_{\mathrm{max}}\\ \delta_1 \neq \delta^*_1}}^{i_{\mathrm{max}}} P(\delta^*_1) \sum_{z_1\geq y}\bigg( \frac{z_1-y}{\Delta}+1\bigg)P(Z^*(\delta_1)=z_1) \nonumber \\
    &+C'_1 \!\!\!\sum_{\substack{\delta^*_1, \delta_1=-d_{\mathrm{max}}\\ \delta_1 \neq \delta^*_1}}^{i_{\mathrm{max}} } \hspace{-1.6em} P(\delta^*_1) \! \! \sum_{z_1\geq y}\!\!\! \bigg(\!\frac{z_1-y}{\Delta}\!+\!1\! \bigg)\!P(Z^*_1(\delta_1)=z_1)2^{\sigma_1(z_1-y)} \nonumber \\
    &- \sum_{\substack{\delta^*_1, \delta_1=-d_{\mathrm{max}}\\ \delta_1 \neq \delta^*_1}}^{i_{\mathrm{max}} } \hspace{-1em} P(\delta^*_1) \sum_{z_1\geq y} \! P(Z^*_1(\delta_1)=z_1) \frac{1-\Delta^{-1}(y-z_1) }{1-2^{-b'}} \nonumber \\
    &=-\frac{2^{-b'}}{1-2^{-b'}} \sum_{\substack{\delta_1^*, \delta_1=-d_{\mathrm{max}}\\ \delta_1\neq \delta^*_1}}^{i_{\mathrm{max}}} P(\delta^*_1) \sum_{z_1\geq y}  \frac{z_1-y}{\Delta}P(Z^*_1(\delta_1)=z_1) \nonumber \\
    &  -\frac{2^{-b'}}{1-2^{-b'}} \sum_{\substack{\delta_1^*, \delta_1=-d_{\mathrm{max}}\\ \delta_1\neq \delta^*_1}}^{i_{\mathrm{max}}} P(\delta^*_1) \sum_{z_1\geq y} P(Z^*_1(\delta_1)=z_1)    \nonumber \\
    &+C'_1 \sum_{\substack{\delta^*_1,\delta_1=-d_{\mathrm{max}}\\ \delta_1 \neq \delta^*_1}}^{i_{\mathrm{max}}} P(\delta^*_1) \sum_{z_1\geq y} P(Z^*_1(\delta_1)=z_1) 2^{\sigma_1(z_1-y)} \nonumber\\
    &\leq C'_1 \sum_{\substack{\delta^*_1,\delta_1=-d_{\mathrm{max}}\\ \delta_1 \neq \delta^*_1}}^{i_{\mathrm{max}}} P(\delta^*_1) \sum_{z_1\geq y} P(Z^*_1(\delta_1)=z_1) 2^{\sigma_1(z_1-y)} \nonumber \\
    &= C'_1 2^{-\sigma_1 y} \sum_{\delta^*_1, \delta_1=-d_{\mathrm{max}}}^{i_{\mathrm{max}}} P(\delta^*_1) \sum_{z_1\geq y} P(Z^*_1(\delta_1)=z_1)2^{\sigma_1 z_1} \nonumber \\
    &-C'_1 2^{-\sigma_1 y} \!\!\! \sum_{\delta^*_1=-d_{\mathrm{max}}}^{i_{\mathrm{max}}} \!\!\!P(\delta^*_1) \sum_{z_1\geq y} P(Z_1^*(\delta^*_1)=z_1)2^{\sigma_1 z_1}. \label{eq::tau0_star}
\end{align}

Analogous to \eref{eq::pzpp}, the latter term is compressed by using $P(\zeta^*=z)=\sum_{\delta=-d_{\mathrm{max}}}^{i_{\mathrm{max}}} P(\delta)P(Z^*(\delta)=z)$. This is essentially the probability distribution of branch metrics along the correct path, as $Z^*(\delta)$ accounts for any transmission errors for a specific drift change $\delta$, while $P(\delta)$ suitably weights these drift-specific distributions according to the likelihood of net insertions or deletions over a single block. Thus, \eref{eq::tau0_star} becomes
\begin{align}
    &\sum_{j=1}^{\infty} E\bigg[ \sum_{\boldsymbol{v}_0^j \in \tau^*_0} \bigg( \frac{\mu(\boldsymbol{v}_0^j)-y}{\Delta}+1\bigg)\phi(\boldsymbol{v}_0^j) \bigg| T^*_{\mathrm{min}}=y \bigg] \nonumber \\
    &\leq C'_1 2^{-\sigma_1 y} \bigg( \sum_{\delta_1=-d_{\mathrm{max}}}^{i_{\mathrm{max}}} \sum_{z_1\geq y} P(Z^*_1(\delta_1)=z_1)2^{\sigma_1 z_1} \nonumber\\
    &\qquad \qquad \qquad-\sum_{z_1\geq y} P(\zeta^*_1=z_1)2^{\sigma_1 z_1}\bigg) \nonumber \\
    &\leq C'_1 2^{-\sigma_1 y} \bigg( \sum_{\delta_1=-d_{\mathrm{max}}}^{i_{\mathrm{max}}} g_{0,\delta} (\sigma_1)-\sum_{z_1\geq y} P(\zeta^*_1=z_1)2^{\sigma_1 z_1}\bigg), \label{eq::tao_star_almost}
\end{align}
where $g_{0,\delta}(\sigma)$ refers to the moment generating function of the drift-specific distribution of branch metrics along the correct path, $P(Z^*(\delta)=z)$. Clearly, given the channel parameters, $\sum_{\delta_1=-d_{\mathrm{max}}}^{i_{\mathrm{max}}} g_{0,\delta_1}(\sigma_1)$ is a constant. We can also deal with the second term in \eref{eq::tao_star_almost} by making use of $z^*_\mathrm{min}$, which denotes the minimum value that $\zeta^*$ can assume. Hence,
\begin{align}
    &\sum_{z_1 \geq y} P(\zeta^*_1=z_1)2^{2^{\sigma_1 z_1}} \geq 2^{\sigma_1 z^*_{\mathrm{min}}} \sum_{z_1\geq y} P(\zeta^*_1=z_1) \nonumber \\
    &=2^{\sigma_1 z^*_{\mathrm{min}}} P(\zeta^*_1\geq y) = 2^{\sigma_1 z^*_{\mathrm{min}}} P(2^{\sigma_0 \zeta^*_1} \leq 2^{\sigma_0 y}) \nonumber \\
    &= 2^{\sigma_1 z^*_{\mathrm{min}}}(1- P(2^{\sigma_0 \zeta^*_1} > 2^{\sigma_0 y})) \geq 2^{\sigma_1 z^*_{\mathrm{min}}}(1-2^{-\sigma_0 y}g_0(\sigma_0)) \nonumber \\
    &\geq 2^{\sigma_1 z^*_{\mathrm{min}}}(1-2^{-\sigma_0 y}), \nonumber
\end{align}
where $\sigma_0$, as first introduced in \eref{eq::sig0_int}, is negative and satisfies $g_0(\sigma_0)=1$. Upon incorporating this into \eref{eq::tao_star_almost}, we infer that
\begin{align}
    &\sum_{j=1}^{\infty}E\bigg[\sum_{\boldsymbol{v}_0^j \in \tau^*_0} \bigg(\frac{\mu(\boldsymbol{v}_0^j)-y}{\Delta}+1 \bigg) \phi(\boldsymbol{v}_0^j) \bigg| T^*_{\mathrm{min}}=y\bigg] \nonumber \\
    &\leq (\hspace{-2.5mm}\sum_{\delta_1=-d_{\mathrm{max}}}^{i_{\mathrm{max}}} \hspace{-3mm} g_{0,\delta_1} (\sigma_1) -2^{\sigma_1z^*_{\mathrm{min}}} )C'_1 2^{-\sigma_1 y}+C'_12^{\sigma_1 z^*_{\mathrm{min}}-(\sigma_0+\sigma_1)y}. \label{eq::tau0_star_final}
\end{align}

Finally, we combine \eref{eq::ec1t_3}, \eref{eq::UB_tau0p} and \eref{eq::tau0_star_final} to establish the following bound on the average decoding effort per block.
\begin{align}
& E[C(\boldsymbol{v}^*_0) |T_\mathrm{min}^*=y]  \leq \sum_{j=1}^{\infty} E\bigg[ \sum_{\boldsymbol{v}_0^j \in \tau'_0} C'(\boldsymbol{v}_0^j) | T^*_{\mathrm{min}}=y \bigg]  \nonumber \\
&\quad +\sum_{j=1}^{\infty} E\bigg[ \sum_{\boldsymbol{v}_0^j \in \tau^*_0} C'(\boldsymbol{v}_0^j) | T^*_{\mathrm{min}}=y \bigg] -\frac{y}{\Delta}+1 \nonumber \\
&\leq   C_12^{-\sigma_1y}+C_22^{-(\sigma_0+\sigma_1)y}+C_3\bigg(-\frac{y}{\Delta}+1\bigg), 
\end{align}

where $C_1$, $C_2$ and $C_3$ are constants for specific channel and convolutional code parameters, and are given by
\begin{align}
    C_1&=\Big(1-2^{-b}+\sum_{\delta_1=-d_{\mathrm{max}}}^{i_{\mathrm{max}}} g_{0,\delta_1}(\sigma_1)-2^{\sigma_1 z^*_{\mathrm{min}}} \Big)C'_1, \nonumber \\
    C_2&=C'_12^{\sigma_1 z^*_{\mathrm{min}}}, \qquad C_3= \frac{2^{-b}-2^{-b'}}{1-2^{-b'}}. \nonumber
\end{align}
\end{document}